\documentclass[twocolumn]{aastex631}

\usepackage{array}
\usepackage{tabularx}

\usepackage{float}
\graphicspath{{./}{figures/}}

\shorttitle{IGRINS observations of WASP-127 b}
\shortauthors{Kanumalla et al.}

\begin{document}

\title{IGRINS observations of WASP-127 b: H$_2$O, CO, and super-Solar atmospheric metallicity in the inflated sub-Saturn}

\author[0009-0005-4890-3326]{Krishna Kanumalla}
\affiliation{School of Earth and Space Exploration, Arizona State University, 781 Terrace Mall, Tempe, AZ, 85287, USA}

\author[0000-0002-2338-476X]{Michael R. Line}
\affiliation{School of Earth and Space Exploration, Arizona State University, 781 Terrace Mall, Tempe, AZ, 85287, USA}

\author[0000-0003-4241-7413]{Megan Weiner Mansfield}
\altaffiliation{NHFP Sagan Fellow}
\affiliation{Department of Astronomy and Steward Observatory, University of Arizona, Tucson, AZ 85719, USA}

\author[0000-0003-0156-4564]{Luis Welbanks}
\altaffiliation{NHFP Sagan Fellow}
\affiliation{School of Earth and Space Exploration, Arizona State University, 781 Terrace Mall, Tempe, AZ, 85287, USA}

\author[0000-0002-9946-5259]{Peter C.B. Smith}
\affiliation{School of Earth and Space Exploration, Arizona State University, 781 Terrace Mall, Tempe, AZ, 85287, USA}

\author[0000-0003-4733-6532]{Jacob L. Bean}
\affiliation{Department of Astronomy \& Astrophysics, University of Chicago, Chicago, IL, USA}

\author[0000-0002-1321-8856]{Lorenzo Pino}
\affiliation{INAF, Astrophysical Observatory of Arcetri, Largo Enrico Fermi 5, I - 50125, Firenze, Italy}

\author[0000-0002-7704-0153]{Matteo Brogi}
\affiliation{Department of Physics, University of Turin, Via Pietro Giuria 1, I-10125, Turin, Italy}
\affiliation{INAF-Osservatorio Astrofisico di Torino, Via Osservatorio 20, I-10025 Pino Torinese, Italy}

\author[0000-0002-2513-4465]{Vatsal Panwar}
\affiliation{Department of Physics, University of Warwick, Coventry CV4
7AL, UK.}
\affiliation{Centre for Exoplanets and Habitability, University of Warwick,
Coventry CV4 7AL, UK.}



\begin{abstract}


High resolution spectroscopy of exoplanet atmospheres provides insights into their composition and dynamics from the resolved line shape and depth of thousands of spectral lines.  WASP-127 b is an extremely inflated sub-Saturn (R$_\mathrm{p}$= 1.311 R$_\mathrm{Jup}$, M$_\mathrm{p}$= 0.16 M$_\mathrm{Jup}$) with previously reported detections of H$_2$O and CO$_2$. However, the seeming absence of the primary carbon reservoir expected at WASP-127 b temperatures (T$_{eq}$ $\sim$ 1400 K) from chemical equilibrium, CO, posed a mystery. In this manuscript, we present the analysis of high resolution observations of WASP-127 b with the Immersion GRating INfrared Spectrometer (IGRINS) on Gemini South. We confirm the presence of H$_2$O (8.67 $\sigma$) and report the detection of CO (4.34 $\sigma$). Additionally, we conduct a suite of Bayesian retrieval analyses covering a hierarchy of model complexity and self-consistency. When freely fitting for the molecular gas volume mixing ratios, we obtain super-solar metal enrichment for H$_2$O abundance of log$_{10}$X$_\mathrm{H_2O}$ = --1.23$^{+0.29}_{-0.49}$ and a lower limit on the CO abundance of log$_{10}$X$_\mathrm{CO}$ $\ge$ --2.20 at 2$\sigma$ confidence. We also report a tentative evidence of photochemistry in WASP-127 b based upon the indicative depletion of H$_2$S. This is also supported by the data preferring models with photochemistry over free-chemistry and thermochemistry. The overall analysis implies a super-solar ($\sim$ 39$\times$ Solar; [M/H] = $1.59^{+0.30}_{-0.30}$) metallicity for the atmosphere of WASP-127 b and an upper limit on its atmospheric C/O ratio as $<$ 0.68.

\end{abstract}



\section{Introduction} \label{sec:intro}

Over the past two decades, space-based low-to-moderate resolution spectroscopy (\textit{R} $\sim$ 50 - 3000) has been the leading technique in characterizing exoplanet atmospheres. The Hubble (HST), Spitzer, and recently, James Webb (JWST) space telescopes have been successful in detecting several key carbon- and oxygen-bearing molecules in the atmospheres of exoplanets \citep[e.g.,][]{madhu_review_2019, co2_jwst, lili_jwst_ers_nirspec,  welbanks_w107b_nature}. Abundance estimation of these molecules is critical in (i) obtaining the atmospheric metallicity estimate and (ii) constraining the formation location of exoplanets in the protoplanetary disk via the C/O ratio \citep[][]{oberg2011, mordasini2016}.

In the era of JWST, ground-based high spectral resolution (R $>$ 15,000) observations complement the space based observations by enabling velocity resolved information. This is critical to unambiguously identify trace absorbers \citep[e.g.,][]{line2021, pelletier_VO_HRS}, disentangle atmospheric dynamics \citep[e.g.,][]{Ehrenreich_wasp76b_Fe_kink, gandhi2023_hrs_dynamics, nortmann-w127b-crires+}, and probe a wide range of atmospheric pressures/altitudes \citep[e.g.,][]{mrkempton2012, bro16}. Specifically, high-resolution cross-correlation spectroscopy (HRCCS) leverages the time-resolved planetary Doppler motion around the host star to separate the planetary signal from the dominant (stellar and telluric) contaminants \citep[e.g.,][]{sne10, dek13, giacobbe2021}. Recent works \citep[e.g.,][]{peter_wasp77Ab_igrins, brogi23} have shown that ground-base HRCCS observations of ultra-hot Jupiters can provide abundance constraints comparable to or even exceeding those achievable with JWST with a similar observing time. 

We applied these HRCCS methods to analyze a single transit observation of an extremely inflated (R$_\mathrm{p}$ = 1.31 R$_\mathrm{Jup}$) warm sub-Saturn (M$_\mathrm{p}=$ 0.16 M$_\mathrm{Jup}$, $\rho \sim$ 0.09 $\mathrm{g/cm^3}$), WASP-127 b \citep{lam2017}. It orbits a bright ($V$ $\sim$ 10.2, $K$ $\sim$ 8.64) and photometrically quiet G5 type star in a close orbit of 4.17 days. With its calculated equilibrium temperature (T$_\mathrm{eq}$ $\sim$ 1400K) and low gravity ($\log$$g$ $\sim$ 2.14 m/s$^2$), it has an atmosphere with one of the largest estimated scale heights ($\sim$ 2350 km) making it highly amenable for transit spectroscopy.  




Previous transmission spectroscopy observations of WASP-127 b have revealed a slew of absorption features. Using GTC/OSIRIS, \cite{chen2018} detected spectral features from Na, Li, K with hints of H$_2$O absorption in the atmosphere of WASP-127 b. The presence of a gray absorbing cloud deck between a pressure levels of $\sim$ 0.3 and 0.5 mbar was also measured by \cite{allart2020}, whose value is consistent with the results of \cite{Skaf_FeH_HST}. Combining the HST (WFC3, STIS) and Spitzer data, \cite{spake2021} did not detect the Li and K features but reported strong absorption from H$_2$O and CO$_2$. From the single photometric data point at 4.5 $\mu$m, \cite{spake2021} also reported that they were not able to disentangle the role of CO$_2$ and CO in the carbon enrichment of WASP-127 b. As a consequence, the retrieved values of C/O ratio range from sub- to super-solar. The reported absence of CO presents a mystery in the atmosphere of WASP-127 b as there is no thermochemical mechanism that would enrich CO$_2$ content while depleting CO \citep[e.g.,][]{moses2011}.

 Using the high-resolution SPIRou spectrograph ($\sim$ 0.95 - 2.50 $\mu$m; \textit{R} = 70,000) on 3.6 m CFHT, \cite{boucher2023} confirmed the presence of H$_{2}$O and a tentative signal of OH absorption. However, they did not detect CO resulting in a sub-solar C/O ratio from their model fits. Recently, \cite{nortmann-w127b-crires+} reported the detection of CO in WASP-127 b using the CRIRES+ ($\sim$ 1.97 - 2.45 $\mu$m; \textit{R} $\sim$ 140,000) on 8.2 m VLT. Currently, these two high-resolution studies present a contrasting picture on the presence of CO in WASP-127 b. Additionally, they reported a strong $\sim$ 7.7 km/sec equatorial jet in WASP-127 b's atmosphere based on two signals from the morning and evening sides of WASP-127 b's terminator. Based on their H$_2$O detection, the results of \cite{boucher2023} support the signal detection from the blue-shifted evening terminator. However, \cite{boucher2023} do not find a red-shifted signal originating from the morning terminator.

In this manuscript, we present our analysis of the time-resolved IGRINS transit observations of WASP-127 b confirming the presence of H$_2$O and CO molecules. In \S \ref{sec:observations}, we describe our observations and the data reduction. In \S \ref{sec:xcorr_detections}, we explain the atmospheric model description followed by the cross-correlation analysis used to detect molecular signals in WASP-127 b. We detail the retrieval framework in \S \ref{sec:methods_retrievals} and we report our constraints on the atmospheric structure. We discuss the implication of our results in \S \ref{sec:discussion} and we provide conclusions of this work in \S \ref{sec:conclusions}.

\section{Observations and Data Reduction} \label{sec:observations}

Observations of WASP-127 b were taken using the IGRINS spectrograph \citep[\textit{R} $\sim$ 45,000,][]{park-igrins-cite} on the Gemini South 8-m telescope as a part of GS-2021A-LP-107 program (PI: Megan Mansfield). The planet was observed in transmission for a continuous sequence of 4.78 hrs covering a full transit, including 12 min of out-of-transit baseline on each side. The observations were taken in ABBA nodding pattern which resulted in 51 AB pairs for the duration of the observation. The integration time for a single AB pair was 120 sec which resulted in a median SNR of 214 and 196 in H- and K-bands respectively.


\begin{figure}[!ht]
\centering
\includegraphics[scale = 0.4]{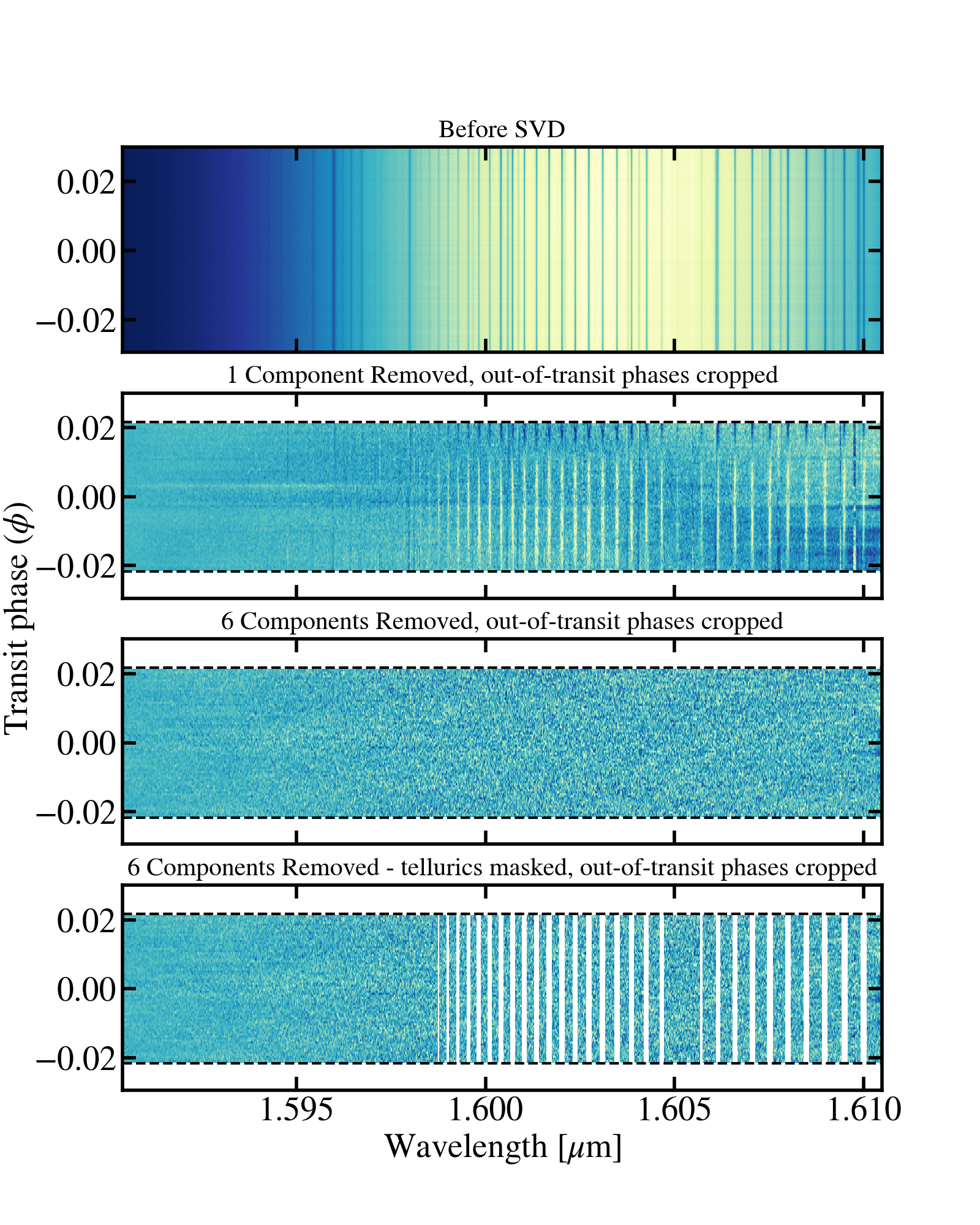}
\caption{An example of the removal of telluric/stellar contamination on one order. The top panels show the ``raw'' flux (counts) with phase (/frame) as a function of wavelength. The second panel is the data matrix after the removal of one eigen component (zeroing out the first singular value) . The out-of-transit frames have been cropped and it can also be seen that the broad-band flux variations due to airmass and throughput variations have been removed. The horizontal black dashed lines represent the beginning and end of transit. The thrid panel shows the matrix after zeroing out the first six components, which largely removes the stellar and telluric features. However, to ensure effective telluric removal, the fourth panel shows strong telluric wavelengths masked from the residual data matrix. The planet's signal is buried inside this noisy matrix. 
\label{fig:svd_test}}
\end{figure}

The IGRINS Pipeline Package \citep[PLP,][]{igrinsplp_lee2016, igrinsplp_mace2018} was used to reduce, optimally extract the spectra, and perform initial wavelength calibrations. As done in \cite{line2021} $\&$ \cite{brogi23}, we performed an additional wavelength adjustment by applying a linear stretch and shift to each spectrum to match the last spectrum in the sequence (closest in time to the PLP-based wavelength calibration source)--precise wavelength alingment with time is critical for the telluric de-trending. We discarded 16 spectral orders due to heavy telluric contamination (median atmospheric transmittance $<$0.7, low SNR)  as well as trim 100 pixels from either side of each order to due to the low throughput. As a result of these steps,  the final un-calibrated flux/counts data cube is of shape: 38 (orders) × 101 (phases [$\phi$]) × 1,848 (pixels/wavelengths [$\lambda$] per order).


The telluric and stellar signals need to be removed from the data as the planetary signal is orders of magnitude smaller. We used the singular value decomposition (SVD) technique \citep{ dek13, bir13} to remove quasi-stationary telluric and stellar spectral features in each each spectral order. The {\tt numpy.linalg.svd} function was used to determine the Eigen vectors (principal components) and the Eigen values (relative contribution of each vector) of the $N_{\phi}$ $\times$ $N_{\lambda}$ matrix (applied to the full sequence including out of transit frames). For all the matrices, we saved the SVD output as two reconstructed matrices: one matrix constructed using the first $N_c$ lower order principal components (i.e., the scaling matrix) and another matrix constructed from the higher order ($>$ $N_c$) principal components (i.e., the residual matrix) to generate the scaling and residual data cubes. The planet signal is contained in this residual data cube buried within the post-SVD residual noise. The resulting products of SVD on one of the orders of WASP-127 b's data is shown in Figure \ref{fig:svd_test}. For our analysis, we used six components for all orders as it resulted in the maximum detection SNR for H$_2$O and CO molecules (see \S \ref{subsec:methods_xcorr}). However, as discussed in \cite{brogi23}, we found that our abundance constraints are weakly dependent on the number of components used.  

To reduce noise in the cross-correlation analysis from the out-of-transit frames where we do not expect any planetary signal, we cropped 15 frames before ingress and 9 frames after egress from the residual data cube post-SVD. Additionally, we masked the wavelengths of the residual data cube where the telluric features are dominant (atmospheric transmittance $<$ 0.90) using the Earth's telluric template obtained from ESO's SkyCalc\footnote{https://www.eso.org/observing/etc/skycalc/}.This helps mitigate any additional contamination in the residual data due to imperfect telluric removal with SVD. The combined effect of these two steps provided a significant increment to our CO signal-to-noise ratio (SNR) in the CC maps (Figure \ref{fig:appendix-flowchart})

\section{Molecular Signal Detection \& Analysis} \label{sec:xcorr_detections}

An important step in any HRCCS analysis is to identify the presence of absorbers within the exoplanet atmosphere. This provides a zeroth order estimate of the nature of the planetary atmosphere. However, because the individual absorption lines are weak compared to the noise level, to identify the atmospheric features, a model template must be cross-correlated as a function of velocity with the residual data (bottom panel in Figure \ref{fig:svd_test}). We first discuss the model template setup and then the cross-correlation analysis.  

\subsection{Model Template Description}
\label{subsec:model_templates}

We first generated a 1D-radiative convective thermochemical equilibrium (1D-RCTE)  atmosphere using {\tt ScCHIMERA} tool \citep{piskorez_2018_rcte, mansfield2021}. ScCHIMERA produces a converged atmospheric structure i.e., gas volume mixing ratios (VMRs) and temperature with pressure (TP profile) given the following inputs: the incident stellar flux \citep[a Phoenix stellar model,][]{phoenix-models-huesser-2013}, internal/effective temperature \citep[taken to be 520K based upon ][]{thorngren-fortney-2019}, planetary/stellar system properties \citep{lam2017}, and the elemental abundances (scaled from \cite{Lodders2009} with an atmospheric metallicity [M/H] and a C/O ratio). We assumed a solar composition atmosphere with no clouds for the initial analysis. 

We included the line opacities from H$_2$O \citep{polyansky2018exomol}, CO \citep{li-co-xsecs}, CO$_2$ \citep{hitemp2010}, CH$_4$ \citep{hargreaves2020accurate}, C$_2$H$_2$ \citep{polyansky2018exomol}, HCN \citep{barber-hcn-xsecs},  NH$_3$ \citep{coles-nh3-xsecs}, OH \citep{rothman2010hitemp}, H$_2$S \citep{azzam-h2s-xsecs}, and FeH \citep{polyansky2018exomol} as well as H$_2$-H$_2$/-He collision-induced continuum absorption \citep[CIA,][]{karman2019hitran}. Cross-sections were pre-computed using the HELIOS-K tool \citep{heliosk2021}. We chose these species as they are the most plausible species to exist under WASP-127 b conditions \citep{burrows_sharp_1999}. 

The 1D-RCTE profiles were then used to generate a high-resolution (R=250,000) transmission spectrum with the CHIMERA transmission forward model \citep{line2013, kreidberg2015, bell_wasp80b_ch4}--upgraded to run on graphics processing units (GPUs). The spectra were then broadened (assuming a rotation kernel \footnote{Although this does not account for additional deviations from solid body rotation, we expect that the low equatorial vsini of WASP-127 b ($\sim$ 1.63 km/sec) will not heavily alter the line shape within IGRINS resolution element ($\sim$ 6.67 km/sec)}  with a planetary rotation velocity, $v$sin$i$ = 1.63 km/sec) and then convolved with the IGRINS instrumental profile (a Gaussian kernel with a FWHM of 5.5 model pixels). 

\subsection{Cross-Correlation Analysis} \label{subsec:methods_xcorr}

Before cross-correlating the transmission model template with the data, it is necessary to ensure that the linear transformation effects from the SVD operation on the data cube are also applied on the model template. As detailed in \cite{brogiline2019} and \cite{line2021}, we injected the model template into the in-transit frames of scaling matrix. This model-injected data cube can be assumed to contain the ``true'' planet's signal with our model assumptions. We then reapply the SVD on the model-injected data cube to recover the appropriately modified model template that can then be directly cross-correlated with the residual data cube. 

\begin{figure}[!ht]
\begin{center}
\includegraphics[scale = 0.31]{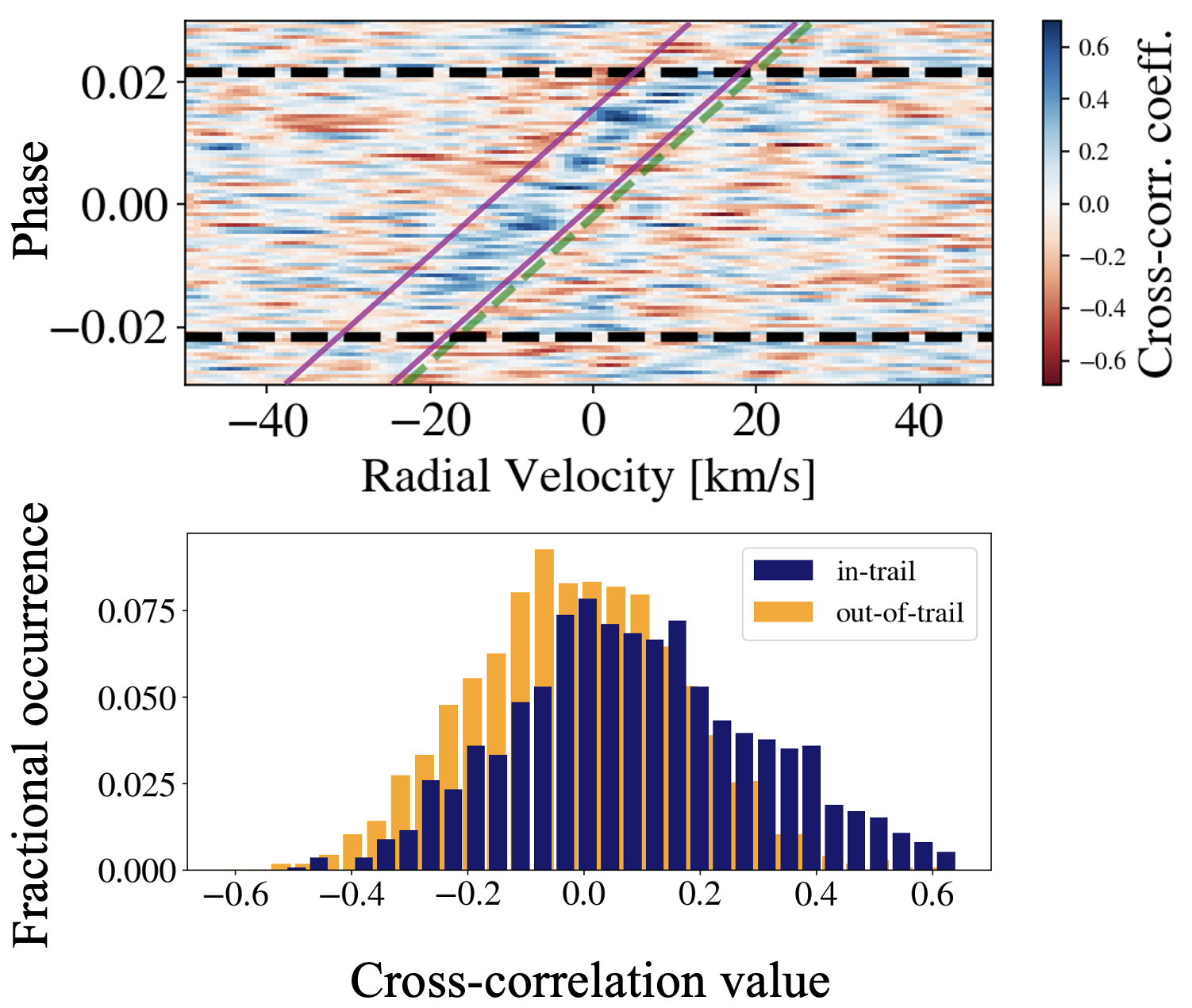}
\caption{ The CC trail of WASP-127 b generated by cross-correlating the residual data with a solar RCTE model with opacity from all the molecular absorbers (i.e., full model). The atmospheric signal can be seen blue shifted ($\sim$ 6 km/sec) from the planet's RV trace (indicated by the green dashed line). The phases during which the planet is in-transit are in between the two dashed horizontal black lines. For the t-test, the in-trail CC values were collected from the region in between the purple lines. The bottom panel shows the normalised/probability histograms of the in-trail and out-of-trail populations. These two histograms show a significant difference,  implying the independance between these two samples. 
\label{fig:trail_hist}}
\end{center}
\end{figure}

For each spectral order, we cross-correlated the residual data matrix with the recovered model template to produce a Pearson cross-correlation value (CC) at each observed phase ($\phi$). Using spline interpolation \citep{bro14}, the forward model was Doppler-shifted based on the planet's velocity ($V_p$) given by,

\begin{equation}
    V_{p}(\phi) = V_{sys} + V_{bary}(t) + K_p \text{sin}(2\pi\phi)
    \label{eq:vp}
\end{equation}

\noindent where, $K_p$ denotes the semi-amplitude of the planet radial velocity, $V_{sys}$ is the star-planet systemic velocity, and $V_{bary}$ denotes the barycentric velocity of the observer. For the WASP-127 system, the literature reported value of systemic velocity is $V_{sys_0}$ = --8.64 $\pm$ 0.89 km/sec \citep{gaia_dr2} and the calculated $K_{p_\mathrm{0}}$ (= K$_{\star}$ $\frac{\mathrm{M_{\star}}}{\mathrm{M_{pl}}}$) is 132.93 $\pm$ 25 km/sec. For a given template model, the cross-correlation function (CC value as a function of velocity) reaches it maximum value at a particular $V_p$ at each phase. The CCF values are summed for all orders and mean-subtracted to generate a trail of positive CC values in $V_{p}$ as a function of $\phi$. 

\begin{figure*}[!ht]
\centering
\includegraphics[scale = 0.3]{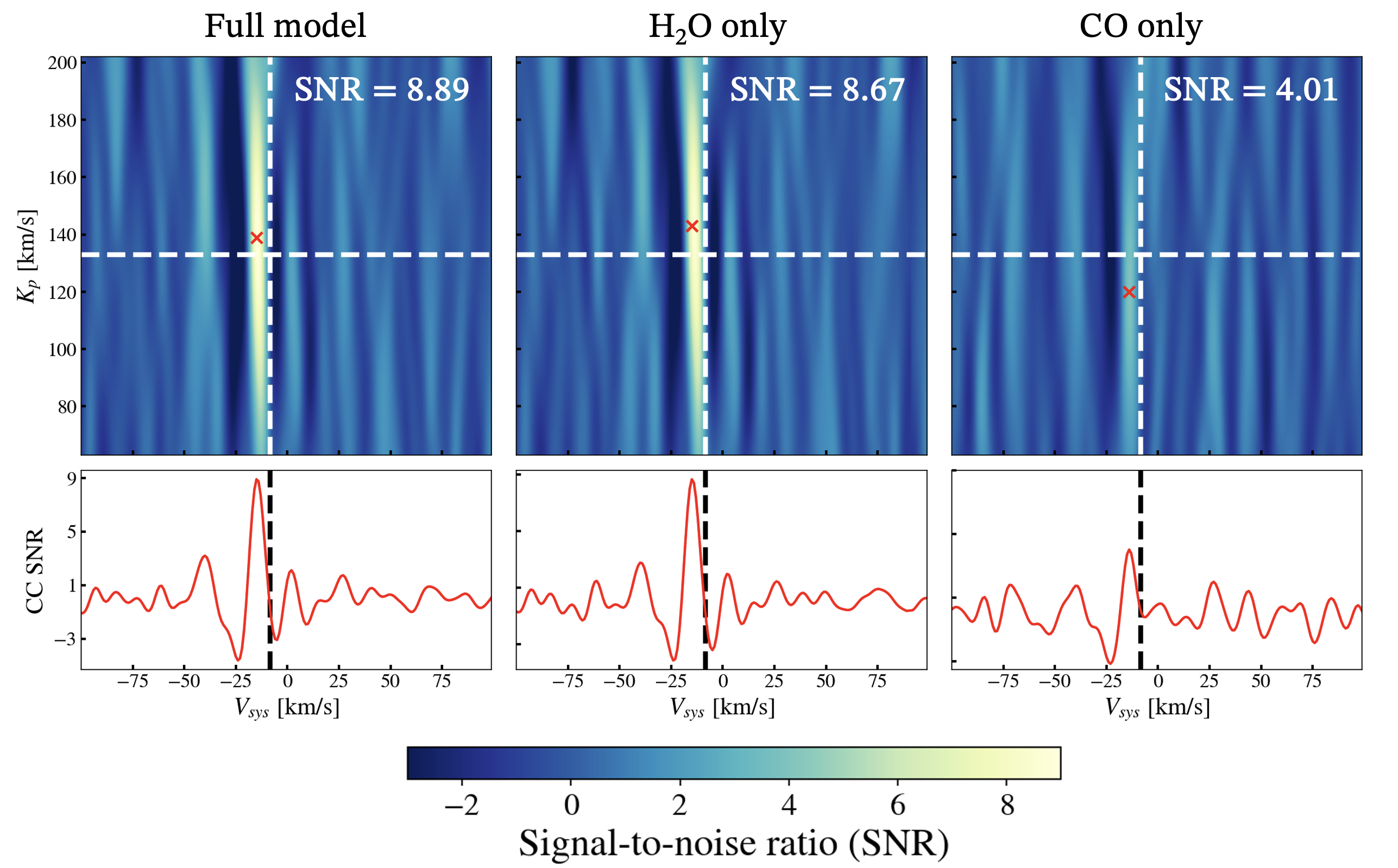}
\caption{
The detections of H$_2$O, CO, and total atmospheric signal in the atmosphere of WASP-127 b. The CC maps shown here are generated by cross-correlating the residual data with solar composition RCTE models with opacity from H$_2$O alone, CO alone, and all absorbers respectively. The tellurics have been masked and the out-of-transit frames have been cropped in the residual data. The white dashed lines in these maps denote the literature reported values of $V_{sys_0}$ and $K_{p_0}$ for the WASP-127 b system. The red cross indicates the peak value of the SNR within each map. Similar to the trail, the peak signal in these maps can be seen blue shifted. The bottom red curves under each panel shows the row/cross-section of the CC SNRs at the peak $K_p$. 
\label{fig:3_ccf_maps}}
\end{figure*}

\begin{figure}[!hb]
\begin{center}
    \includegraphics[scale = 0.25]{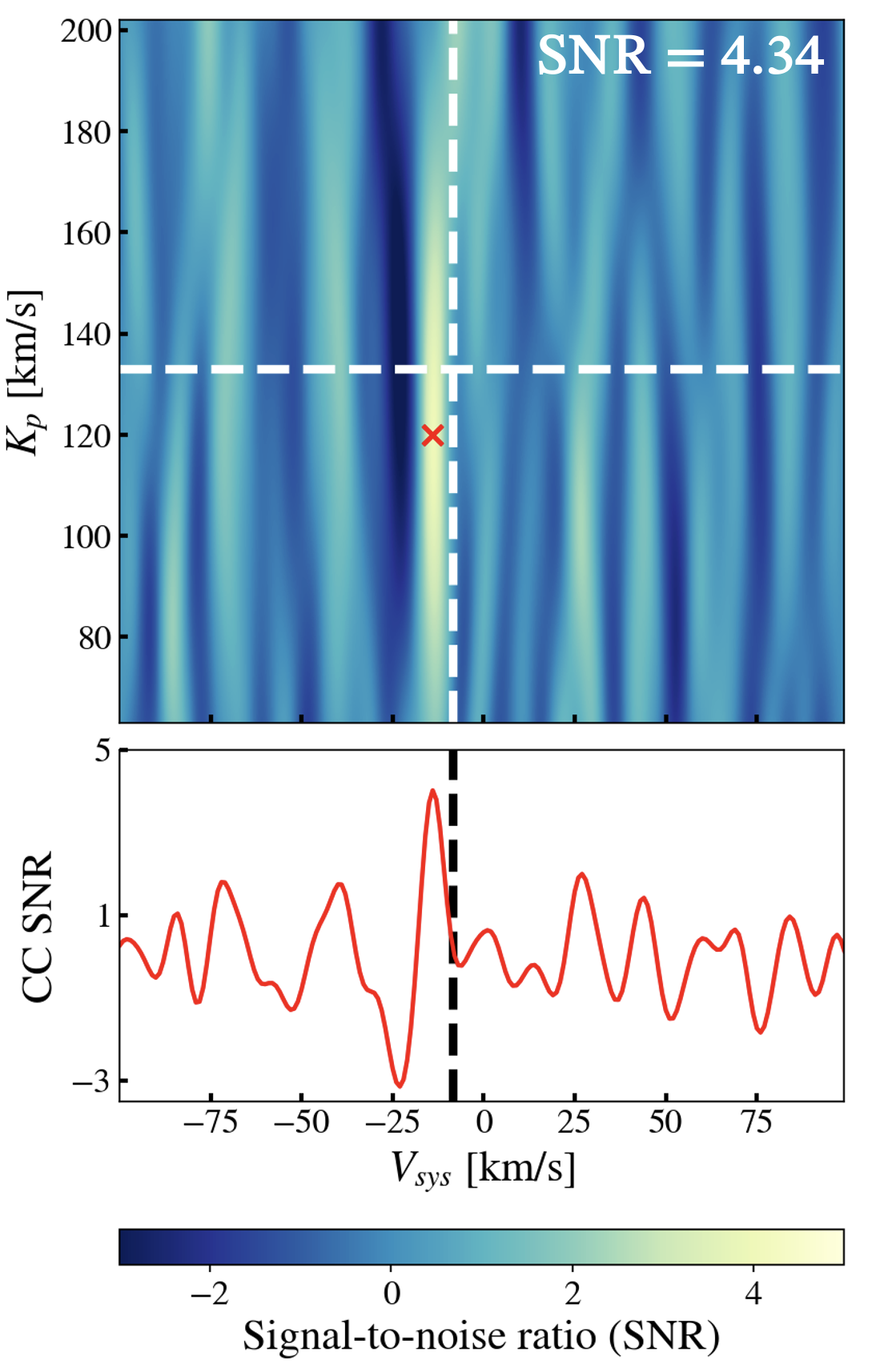}
\caption{The CC map generated by cross-correlating the CO only model with the IGRINS orders where CO has prominent features. By doing so, the CC noise from rest other orders was suppressed and the SNR of CO detection has been increased to 4.34$\sigma$. Similar to Figure \ref{fig:3_ccf_maps}, the white dashed lines indicate the literature $K_{p_\mathrm{0}}$ and $V_{{sys}_\mathrm{0}}$ values. The bottom panel shows the CC cross-section at the peak K$_p$ value.
\label{fig:order_selection_CO}}
\end{center}
\end{figure}

We detected a clear trail of WASP-127 b's atmosphere when the residual data was cross-correlated with a solar composition 1D-RCTE model template containing opacities from all the above mentioned absorbers (termed as the full model). The trail, shown in Figure \ref{fig:trail_hist}, is significantly blue shifted from the planet's RV curve by $\sim$ 6 km/sec. To quantify the detection of planetary signal from this trail, we performed the Welch's t-test between the in-trail and out-of-trail CC values data with the null hypothesis that both of these data are sampled from the same population and therefore are not disparate. From the observed blue shifted atmospheric signal, we generated the in-trail sample using the CC values within $\pm$ 5 km/sec from the observed planetary trail. In Figure \ref{fig:trail_hist}, this is shown as the region in between the purple lines. For the out-of-trail dataset, we selected the CC values that were 35 km/sec away from the trail. From this, the t-test on the in-trail vs out-of-trail CC values confirmed the trail's presence with a 9.24$\sigma$ significance.

Unlike the full model trail, the trails generated using templates of individual molecules are often hard to spot unless the model parameters match well to those of the planet's true signal. Therefore, we produced CC maps over a grid of $K_p$ and $V_{sys}$ values which allow for a clear signal detection. For this, we again cross-correlated the residual data cube with the recovered model template and summed the CCF values for all phases and orders to produce a CC map as seen in Figure \ref{fig:3_ccf_maps}. The SNR in these maps is calculated by dividing the peak CC value by the standard deviation of the whole map.  

\begin{table*}[!ht]
  \centering
  \begin{tabular}{>{\centering\arraybackslash}m{3cm}>{\centering\arraybackslash}m{6cm}>{\centering\arraybackslash}m{6cm}}
    \hline

    \textbf{Parameter} & \textbf{Description} & \textbf{Prior Range} \\
    \hline
    \multicolumn{3}{c}{\textbf{Free Chemistry Retrieval}} \\
    \hline
    
    log$_{10}$$X_{\text{gas}}$ & Logarithmic volume mixing ratios of gases & $\mathcal{U}(-12, 0.3)$ \\
     & (H$_2$O, CO, CO$_2$, CH$_4$, C$_2$H$_2$, HCN,  NH$_3$, OH, H$_2$S, FeH)  & \\
     
    \begin{center}
        $T_{0}$
    \end{center} & Temperature at the top (10$^{-9}$ bars) of atmosphere & $\mathcal{U}(500, 3000)$ K \\
    
    \begin{center}
        log$_{10}$$P_{1,2}$
    \end{center} & Pressure nodes at the top and middle of the inversion layer & $\mathcal{U}(-9, 2)$ bar \\
    
    log$_{10} P_{3}$ & Pressure at the bottom of inversion layer & $\mathcal{U}(-2, 2)$ bar \\
    
    $\alpha_{1,2}$ & Slopes of the TP profile & $\mathcal{U}(0.02, 2)$ K$^{-1/2}$ \\
    
    \hline
    \multicolumn{3}{c}{\textbf{Grid Retrievals}} \\
    
    \hline
    $f$ & Heat redistribution factor & $0.543 \rightarrow 1.08$ (11 grid points) \\
    
    [M/H] & Logarithmic atmospheric metallicity & $-0.25 \rightarrow 2.5$ (21 grid points)\\
    
    C/O & Carbon-to-oxygen ratio & $0.1 \rightarrow 0.95$ (15 grid points)\\
    
    \hline
    \hline
    
    log$_{10}P_{cl}$ & Logarithmic cloud top pressure & $\mathcal{U}(-9, 2)$ bar \\

    $K_{p}$ & Radial velocity semi-amplitude & $\mathcal{U}(102, 182)$ km/sec \\
    
    $V_{sys}$ & Systemic velocity & $\mathcal{U}(-20, 20)$ km/sec \\
    
    log$_{10}P_{\text{ref}}$ & Logarithmic reference pressure at $R_{\text{pl}}$ & $\mathcal{U}(-9, 2)$ bar \\
    
    log $a$ & HRCCS specific scaling factor & $\mathcal{U}(-1, 1)$ \\
    
    $\times {Rp}$ & Scaling factor for the radius of planet & $\mathcal{U}(0.5, 1.5)$\\
    
    \hline
  \end{tabular}
  \caption{The description and priors of the parameters in the retrievals. $\mathcal{U} (a,b)$ denotes a uniform prior from a to b. In the grid retrievals, priors on $f$, [M/H], and C/O are uniform from their lowest to the highest grid point value.}
  \label{table:retrieval_params}
\end{table*}

We obtained a strong detections of H$_2$O (SNR = 8.67$\sigma$) and CO (SNR = 4.01$\sigma$) from the CC maps generated using the model templates containing H$_2$O and CO opacities respectively. These CC maps, along with the one generated using the full model, are shown in the panels of Figure \ref{fig:3_ccf_maps}. We noticed a similar blue shifted $V_{sys}$ offset seen in the CC trails in all the three maps. We also generated the CC maps using the individual model templates of other absorbers (i.e., CH$_4$, C$_2$H$_2$, HCN,  NH$_3$, OH, H$_2$S, FeH, CO$_2$); however, we did not detect any strong signal peaks (Figure \ref{fig:appendix-non-detections}). As a result, the CC map generated using a model with H$_2$O+CO opacity produced an SNR peak that was as significant as the full model signal.

To validate our CO detection, we only summed the CCFs from the spectral orders where the opacity from CO was dominant. These correspond to eight spectral orders in the H and K bands with the ro-vibrational transitions of CO (Figure \ref{fig:appendix-order_selection_models}). By doing so, we boosted the significance of CO detection from a 4.01 $\sigma$ to 4.34 $\sigma$ (Figure \ref{fig:order_selection_CO}).

\section{Atmospheric Retrieval Analysis} \label{sec:methods_retrievals}

In order to obtain quantitative information about the composition, clouds, and the temperature structure, Bayesian inference (i.e., atmospheric retrievals) must be performed. We used the two common retrieval methods based on the free-chemistry \citep[e.g.,][]{madhuseager2009, kreidberg2015} and the grid-based  \citep[e.g.,][]{brogi23, bell_wasp80b_ch4} chemically consistent atmosphere paradigms. The full description and priors for all the parameters in these retrievals are given in Table \ref{table:retrieval_params}. We used the log-likelihood framework from \cite{brogiline2019} and sampled the prior space with the {\tt pymultinest} tool \citep{Feroz2009, Buchner2014}. 

\subsection{Free-chemistry retrieval}
\label{subsec:free-retrieval}

In the free-chemistry retrieval, we fitted for  the constant-with-altitude VMRs for H$_2$O, CO, CO$_2$, CH$_4$, C$_2$H$_2$, HCN,  NH$_3$, OH, H$_2$S, and FeH. We used the six-parameter temperature-pressure (TP) profile description of \cite{madhuseager2009} . We also fitted for an opaque grey cloud-top pressure (log$_{10}$$P_{cl}$) as well as the planetary radius (as a scaling factor to the reported radius, $\times$$R_p$) and the reference pressure (log$_{10}$$P_{ref}$) for that radius. The latter two account for the uncertainty in where to vertically position the hydrostatic grid which impacts both the planetary gravity with altitude and the stretching/shifting to the overall spectrum \citep{welbanks_degeneracies}. We also included a  scale factor (log$_{10}$$a$) to account for uncertainties in the stretching of the planetary signal during SVD. Finally, we also retrieved for both the $K_p$ and $V_{sys}$ parameters. 

\begin{figure*}[h!]
\begin{center}
\includegraphics[scale = 0.3]{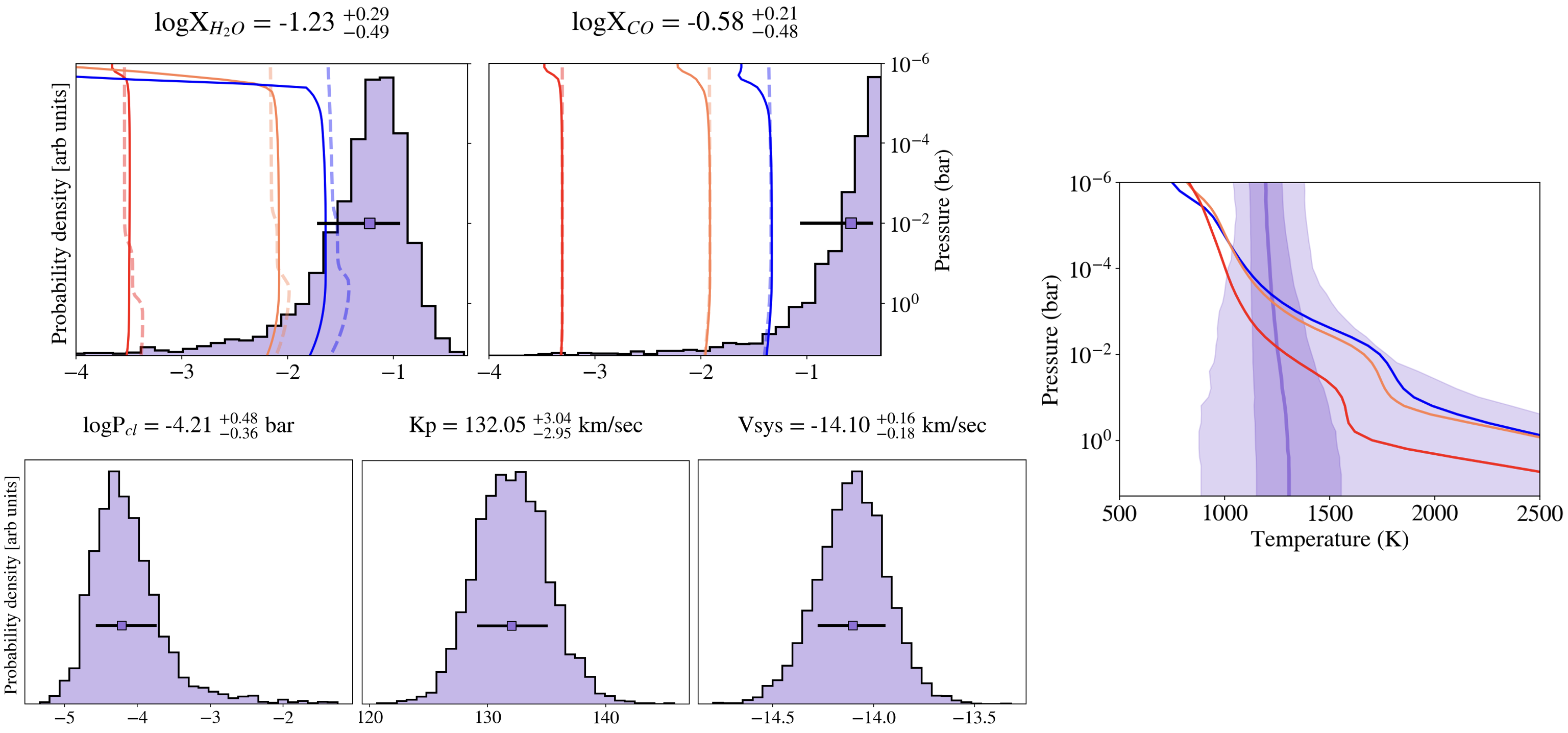}
\caption{The summary of the free-chemistry retrieval showing a subset of five posterior distributions (left) and retrieved TP profile (right) using the \cite{madhuseager2009} parameterisation. In the posterior distributions, the square marker and the horizontal line represent the median value and 1$\sigma$ error bar on each parameter. For comparison, the VMRs of H$_2$O and CO from the RCTE (dashed) and RCPE (solid) models with 1$\times$, 25$\times$, 100$\times$ solar metallicity are also shown. These correspond to red, coral, and blue lines respectively. The retrieved TP profile is shown on the right with lighter shades of purple representing 1$\sigma$ and $2\sigma$ regions. The RCPE TP profiles from 1$\times$ (red), 25$\times$ (coral), 100$\times$ (blue) solar metallicity are also plotted for reference. \label{fig:results-free-summary}}
\end{center}
\end{figure*}

Figure \ref{fig:results-free-summary} summarizes the constraints from the free retrieval. The full posterior is summarized with a corner plot in the Appendix (Figure \ref{fig:appendix-full-free-retrieval}). The free retrieval finds log$_{10}$$X_\mathrm{H_2O}$ = --1.23$^{+0.29}_{-0.49}$ and a lower limit on the CO abundance, log$_{10}$$X_\mathrm{CO}$ $>$ --2.20 at 2$\sigma$. Only upper limits on the remaining 8 gases were obtained, consistent with their non-detections. Additionally, we obtained a bounded constraint on cloud-top-pressure, log$_{10}$$ P_{cl}$ = --4.21$^{+0.48}_{-0.36}$. The temperature profile is largely unconstrained below the cloud deck but is generally monotonically decreasing with altitude.  The obtained $K_p$ and $V_{sys}$ constraints i.e., 132.05$^{+3.04}_{-2.95}$ $\&$ --14.10$^{+0.16}_{-0.18}$ km/sec respectively, are consistent with the peaks from the CC maps. Based on the obtained abundance constraints, we calculated the estimates on [M/H] and C/O \citep[eqs (1) and (2),][]{brogi23} as $>$ 0.07 (2$\sigma$) and $>0.60$ (2$\sigma$) respectively.

\subsection{Grid-based retrievals}
\label{subsec:grid-retrievals}

In the grid-based retrievals, we directly retrieved the [M/H] and a C/O, as well as the heat redistribution factor $f$ based upon a pre-computed grid of  1D-raditive-convective thermochemical/photochemical equilibrium (1D-RCTE/RCPE) model atmospheres \citep[e.g.,][]{brogi23, bell_wasp80b_ch4}. To assess the role of disequilibrium chemistry in influencing the inferred abundances, we performed grid retrievals with both the thermochemical and the photochemical grids to the data. 

To fit the IGRINS observations, we generated on-the-fly transmission spectrum given an interpolated model atmosphere, dictated from the {\tt pymultinest} parameter draws for [M/H], C/O, and $f$. The model atmosphere (gas VMRs and TP profile) for each parameter draw is generated via regular grid interpolation (using {\tt scipy.interpolate.RegularGridInterpolator}) on the 3,465 grid points. To remain consistent with the free retrieval, log$_{10} P_{cl}$, log$_{10} P_{ref}$, $\times R_{p}$, log$_{10} a$, $K_p$, and $V_{sys}$ parameters were also included.


\begin{figure}[!ht]
\begin{center}
\includegraphics[scale = 0.2]{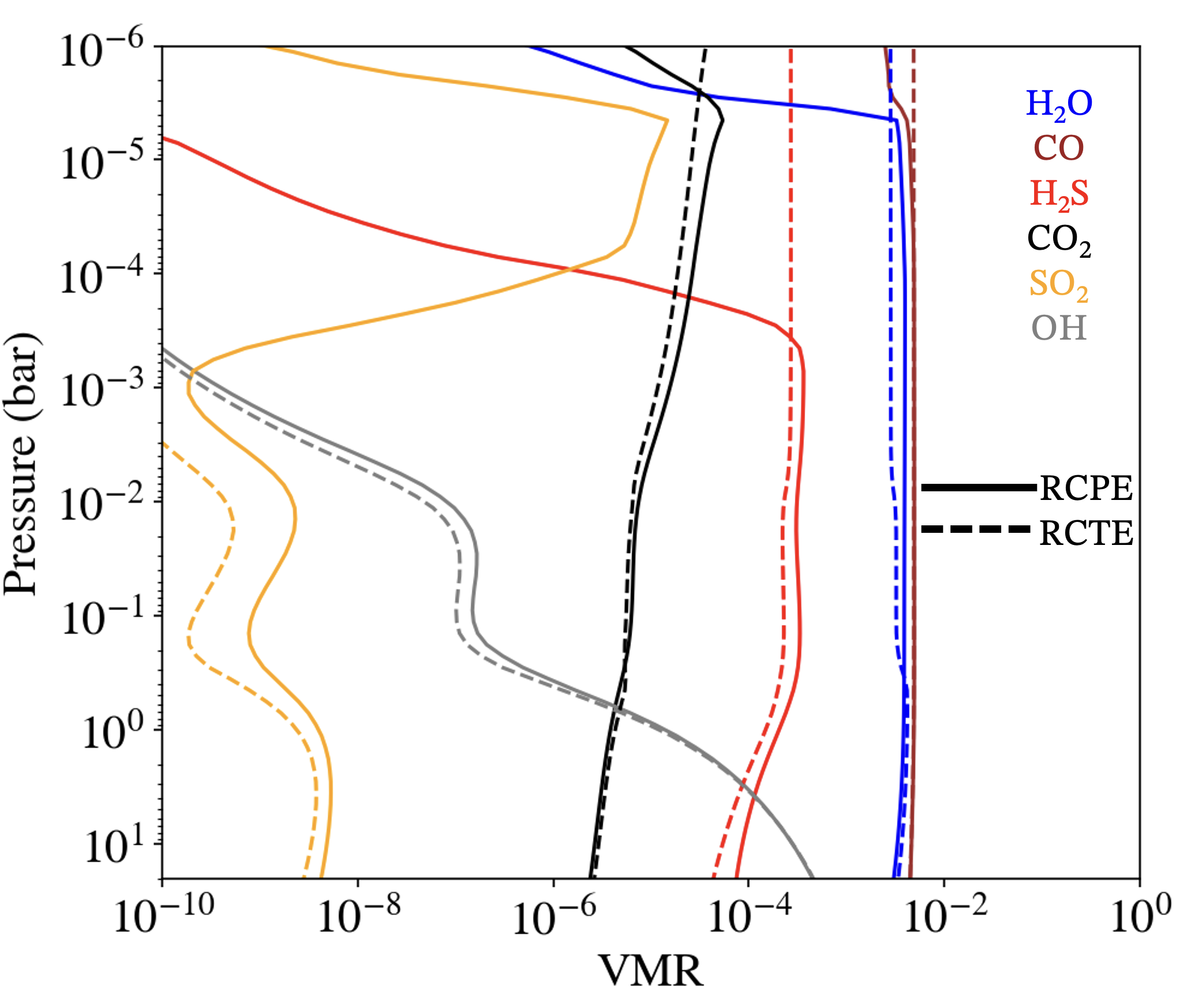}
\caption{VMR profiles of several dominant absorbers in IGRINS coverage. These profiles are calculated for an atmosphere with 10$\times$ solar metallicity with \textit{f} and C/O fixed at 1 and 0.5 respectively. The solid lines indicate the VMR profiles from a RCPE model and the dashed lines indicate the same from a RCTE model. Both H$_2$O and H$_2$S are highly impacted by photochemistry, however H$_2$S is heavily depleted deeper in the atmosphere unlike H$_2$O. 
\label{fig:bf_RCPE_VMRs}}
\end{center}
\end{figure}

\begin{figure*}[ht!]
\begin{center}
\includegraphics[scale = 0.3]{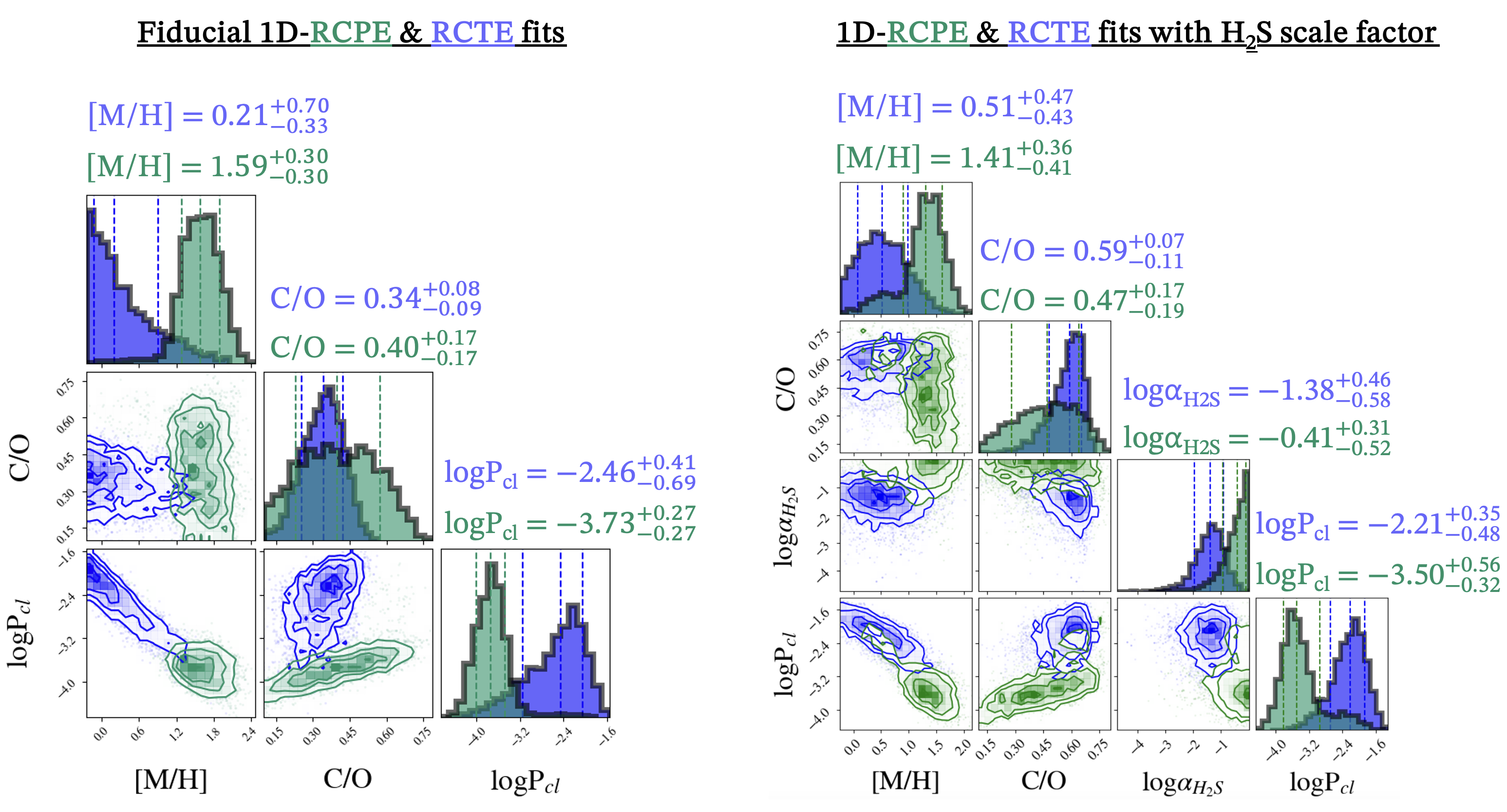}
\caption{The summary of our grid retrievals showing the retrieved posteriors on [M/H], C/O, and log$_{10}$$P_{cl}$ from the fiducial (left) retrieval and the test (right) retrieval. In the right inset, the additional log$_{10}$$\alpha_{H2S}$ parameter corresponds to the scaling factor on the H$_{2}$S abundance (VMR) profile from the test retrievals. In both the left and right insets, the blue contours and marginal distributions are from the thermochemical grid retrieval and the green contours and marginal distributions correspond to the photochemical grid retrievals. The dotted blue and green lines indicate the median (middle dotted line) and 1$\sigma$ confidence regions for a given posterior.
\label{fig:results-grid-summary}}
\end{center}
\end{figure*}

\begin{figure}[ht!]
\begin{center}
\includegraphics[scale = 0.48]{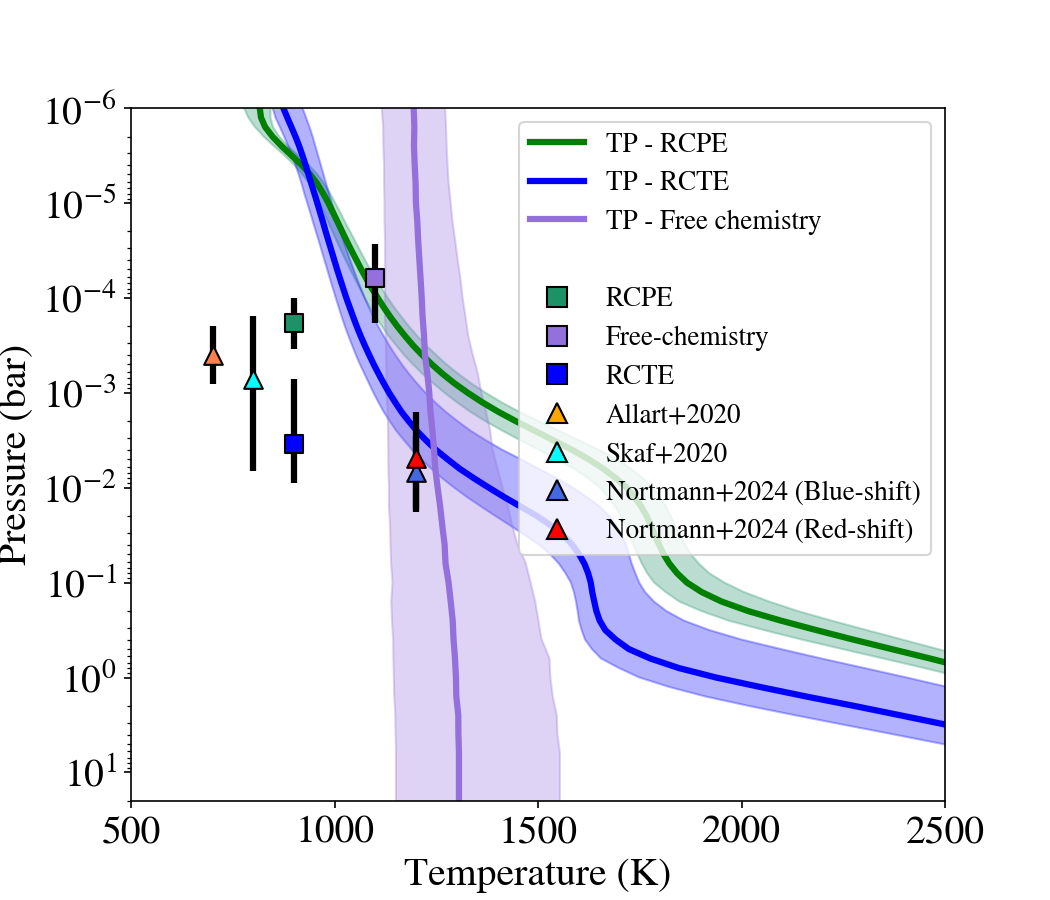}
\caption{The retrieved TP profiles from the RCPE and RCPE grid retrievals. The TP profile from the free-chemistry retrieval is also shown for reference. In these profiles, the lighter shades of the colors represent the 1$\sigma$ confidence region. The constraint on the pressure level of the gray absorbing cloud from all the retrievals (square markers) and previous studies (triangle markers) are shown along with their 1$\sigma$ errors (black bars). Among these, the pressure level of the cloud-deck from the fiducial RCPE retrieval is consistent with the two previous studies, whereas the results from RCTE retrieval are consistent with constraints from \cite{nortmann-w127b-crires+}.
\label{fig:TP-grid_retrievals}}
\end{center}
\end{figure}

Figure \ref{fig:results-grid-summary} shows the results from the grid-retrieval scenarios. Full corner plots are shown in the Appendix (Figure \ref{fig:appendix-both-grid-retrievals}). From these fiducial grid retrieval fits (left corner plot of Figure \ref{fig:results-grid-summary}), we achieved disparate constraints on the [M/H] and C/O with a RCTE (blue) and the RCPE (green) model assumptions. Under the RCTE model assumption, we obtained upper limit on the [M/H] ($<$ 1.64; 2$\sigma$) and a sub-solar C/O (0.34$^{+0.08}_{-0.09}$). However, with a RCPE model assumption, we obtained a bounded constraint on the [M/H] = 1.59$^{+0.30}_{-0.30}$ and an upper limit on the C/O as $<$0.68. The retrieved pressure of the cloud deck also varies between the two retrievals, corresponding to $\sim$ 3.4 mbar (log$_{10}$$ P_{cl}$ = --2.46$^{+0.41}_{-0.69}$) and $\sim$ 0.18 mbar (log$_{10}$$ P_{cl}$ = --3.73$^{+0.27}_{-0.27}$) from the RCTE and RCPE retrievals respectively. The $K_p$ and $V_{sys}$ velocities remain consistent between the two grid retrievals (and with the free retrieval) indicating that neither one of these has converged to an erroneous parameter space. The heat redistribution parameter, although unconstrained, also remains consistent between the two retrievals yielding a similar atmospheric temperature structure for both (Figure \ref{fig:TP-grid_retrievals}).

We further investigated the source of the discrepancy between the RCTE and RCPE retrieval constraints. To do this, we first inspected the model atmosphere structures under the RCPE/RCTE assumptions for a representative composition to understand which species are most affected by disequilibrium. Figure \ref{fig:bf_RCPE_VMRs} shows an example atmospheric configuration for f = 1, [M/H]= 1 and a C/O = 0.5. From this, we noted that H$_2$S is the third most dominant molecular absorber in our models and is strongly influenced by photo-dissociation.

To test the impact of H$_2$S on the inferred composition, we ran two grid retrievals where we scaled the VMR profile of H$_{2}$S with a multiplicative factor (log$_{10}$$\alpha_{H_2 S}$) in both grids. We set a uniform prior ($\mathcal{U}(-6, 0)$) for this factor.  When applied to the RCTE grid (Figure \ref{fig:results-grid-summary}, right panel), $\alpha_{H2S}$ converged to a value well below unity ( log$_{10}$$\alpha_{H_2 S}$ = --1.38$^{+0.46}_{-0.58}$). In contrast, when applied to the RCPE grid, the scale factor constraint runs up against the upper prior bound (i.e., no scaling, $>$ --1.07 at 2$\sigma$). 

This signified that the inclusion of the H$_2$S scale factor had very little overall impact on the derived constraints from the RCPE grid, but a significant impact on the ones derived from the RCTE grid. The inclusion of $\alpha_{H_2 S}$ decreased the metallicity upper limit from $<$ 1.64 (2$\sigma$) to $<$ 1.47 (2$\sigma$) and altered the sub-solar C/O (0.35$^{+0.10}_{-0.10}$) constraint to a super-solar value (0.59$^{+0.07}_{-0.11}$). However, the pressure level of the cloud remained consistent within 1$\sigma$. Most notably, while such modification substantially transformed our RCTE retrieval results, we noticed that all of our constraints remained consistent (within 1 $\sigma$) between the RCPE retrievals (green contours/distributions in Figure \ref{fig:results-grid-summary}). 


To reinforce our inferences drawn from all of our retrievals, we computed the Bayesian evidences ($\mathcal{Z}$) from each retrieval paradigm to determine the model assumption that is most suitable to our data. We obtained the highest evidence from the RCPE grid retrieval, followed by the free-chemistry ($\Delta ln \mathcal{Z} = 4.10$) and RCTE grid retrieval ($\Delta ln \mathcal{Z} = 33.89$) -- suggesting that the model that includes photochemistry is significantly favored compared to the models from the RCTE and free=chemistry retrievals by 8.54$\sigma$ and 3.33$\sigma$ confidence respectively.

\section{Discussion} 
\label{sec:discussion}

\subsection{The search for CO}

Comparing our analysis with \cite{boucher2023}, the most contrasting difference between our studies is the higher SNR of our observations and the higher number of components we needed to remove to achieve a tentative 3.05 $\sigma$ confidence signal in our CCF map of CO template model (see Figure \ref{fig:appendix-flowchart}). To boost the SNR, we had to (i) mask the strong tellurics, (ii) crop the out-of-transit frames (post-SVD) and (iii) select the spectral orders with high CO opacity in CC map generation. The third strategy solidified the signal detection from applying (i) and (ii) and best works for CO and other molecules with banded opacity structure. However, as the SNR of the planet's signal in HRCCS methods relies on the number of lines, applying (iii) to generate CC map with H$_2$O only model produced a signal with a significantly reduced SNR due to less number of H$_2$O lines captured. 


Between the first two strategies, cropping the out-of-transit frames provided a higher boost to the SNR of the CO signal. Recently, studies have shown that higher number of out-of-transit frames assists the SVD technique in efficiently identifying the dominant trends \citep[e.g.,][]{cabot2024, dash2024} in the data. For the same reason, we also performed the SVD on the full dataset including out-of-transit frames. However, with our analysis, we show that cropping the out-of-transit frames post-SVD was highly influential in augmenting the significance of our signal detections. This also bodes well considering that frames without any planet signal cannot contribute to the SNR of the CC map, thus removing them should reduce the noise they pose. However, this statement best applies in the case of SVD detrending. Because, in \cite{dash2024}, it was shown that principal component analysis (PCA) based detrending smears the true exoplanet signal to the out-of-transit portion of the sequence when eigenvalues are re-fitted on the data via multi-linear regression. 

Our CO detection was made possible due to combined effect of the mentioned three strategies. Therefore, we echo the prospect of removing the out-of-transit frames and telluric masking in minimising the contaminants of higher order data components, where we find the planetary signal. In Figure \ref{fig:appendix-flowchart}, we have detailed our steps implemented and their corresponding boost in the CO detection significance.

The consistent position of our signal peaks in our CC maps and in all the retrievals serves a robust indicator that we are finding a consistent atmospheric signal defined by a specific RV curve.  While the deviations from the literature reported $K_p$ value can be attributed to errors on the adapted planetary parameters, the deviations in $V_{sys}$ are best explained only from winds, ephemerides error (e.g., transit mid-point errors), and other 3-dimensional atmospheric effects \citep[e.g.,][]{bro16}. We observed a strong blue shift of $\sim$ 6 km/sec, consistent with the blue shift reported by \cite{boucher2023} based on their H$_2$O detection. However, we do not find any red shifted signals of H$_2$O and CO as seen in \cite{nortmann-w127b-crires+}. This is likely due the comparatively lower spectral resolution offered by IGRINS. Regardless, as discussed in \cite{line2021}, we expect that the V$_{sys}$ offset had a negligible effect on altering our retrieved abundance inferences as we did not observe any correlation with other atmospheric parameters of the retrieval. 

\subsection{Free-chemistry vs Thermochemistry vs Photochemistry}

Altering the H$_{2}$S abundance profile by scaling it produced minimal difference of the results of the RCPE grid retrieval while substantially changing the results of its thermochemical counterpart. This is a direct consequence of the in-built H$_{2}$S depletion at low pressure levels above the cloud deck as a result of photochemistry. Additionally, this also indicated the the inferences drawn from our thermochemical retrieval were greatly influence by H$_{2}$S, a molecule we could not detect both from the CC analysis and the free-chemistry retrieval. 

To further support this reasoning, we assessed the level to which H$_2$S would have been detected, if it was indeed present based on thermochemical equilibrium. This would also verify that our detrending technique and analysis were unbiased in detecting H$_2$S. For this, we removed the best fit fiducial RCPE model from the data and injected the best fit fidicual RCTE model. This simulated dataset containing the best-fit RCTE signal was then cross-correlated with the best fit RCTE model and best fit H$_2$S RCTE models. From Figure \ref{fig:appendix-h2s_injection_recovery}, we were able to recover a signal of H$_2$S with a confidence of 3.08 $\sigma$. 

It is important to acknowledge the biases on estimating [M/H] and C/O ratio from a free-chemistry model assumption. \cite{brogi23} showed that C/O estimates evaluated from the molecular abundance posteriors of a free-chemistry retrieval are overestimated. In their case, it was showed that the abundance estimates of carbon-containing species can exceed the value permissible based on a chemical equilibrium. From our results, we see a similar effect where the obtained constraint on the C/O ratio is $>$0.60 (2$\sigma$), which is primarily derived from our posteriors on X$_\mathrm{H2 O}$ $\&$ X$_\mathrm{CO}$. This estimate would be contradictory given the reported CO$_2$ detection by \cite{spake2021}. We note that, we were only able to place upper limit on the abundance of CO$_2$ from our free-chemistry retrieval. 

RCPE models account for a fundamental process of planetary atmospheres i.e., photochemistry, which is not captured by the models with free-chemistry and thermochemistry. The limitations of free-chemistry retrieval and RCTE grid retrievals made us question the reliability of [M/H] and C/O estimates obtained from them. Based on these reasons and from Bayesian evidence, we quote our final results from our RCPE grid retrieval. 


\subsection{Atmospheric metallicity of WASP-127 b}

With a chemically consistent retrieval, the most recent compositional assessment on WASP-127 b by \cite{nortmann-w127b-crires+} reports solar values for both C/O (0.56$^{+0.05}_{-0.07}$) and atmospheric metallicity ([M/H] = --0.01$^{+0.36}_{-0.39}$). Our RCTE grid retrieval constraints on [M/H] (i.e., 0.21$^{+0.70}_{-0.33}$) and C/O (i.e., 0.34$^{+0.08}_{-0.09}$) are consistent with these values within 1$\sigma$ and 2$\sigma$ confidence regions respectively.

However, a solar metallicity for WASP-127 b would be in tension with the results of \cite{chen2018, Skaf_FeH_HST, spake2021, boucher2023}, which have all reported super-solar metal enrichment. From a retrieval with a free-chemistry assumption, \cite{chen2018}  reported super-solar abundance for logX$_\mathrm{H_2 O}$ as --2.50$^{+0.94}_{-4.56}$.  This result was also supported by \cite{Skaf_FeH_HST} who reported a tighter constraint as logX$_\mathrm{H_2 O}$ $=$ --2.71$^{+0.78}_{-1.05}$. Both the free-chemistry and chemically consistent retrievals of \cite{spake2021} have favoured super-solar abundance, yielding an average of 17 $\pm$ 4$\times$ solar metallicity for H$_2$O, Na, and CO$_2$ under chemical equilibrium. With a free-chemistry assumption, \cite{boucher2023} obtained a solar logX$_\mathrm{H_2 O}$  as --3.0$^{+0.5}_{-0.6}$, however, a super-solar value of logX$_\mathrm{C O_2}$ as --3.7$^{+0.8}_{-0.6}$. 

\begin{figure}
\centering
\includegraphics[scale = 0.4]
{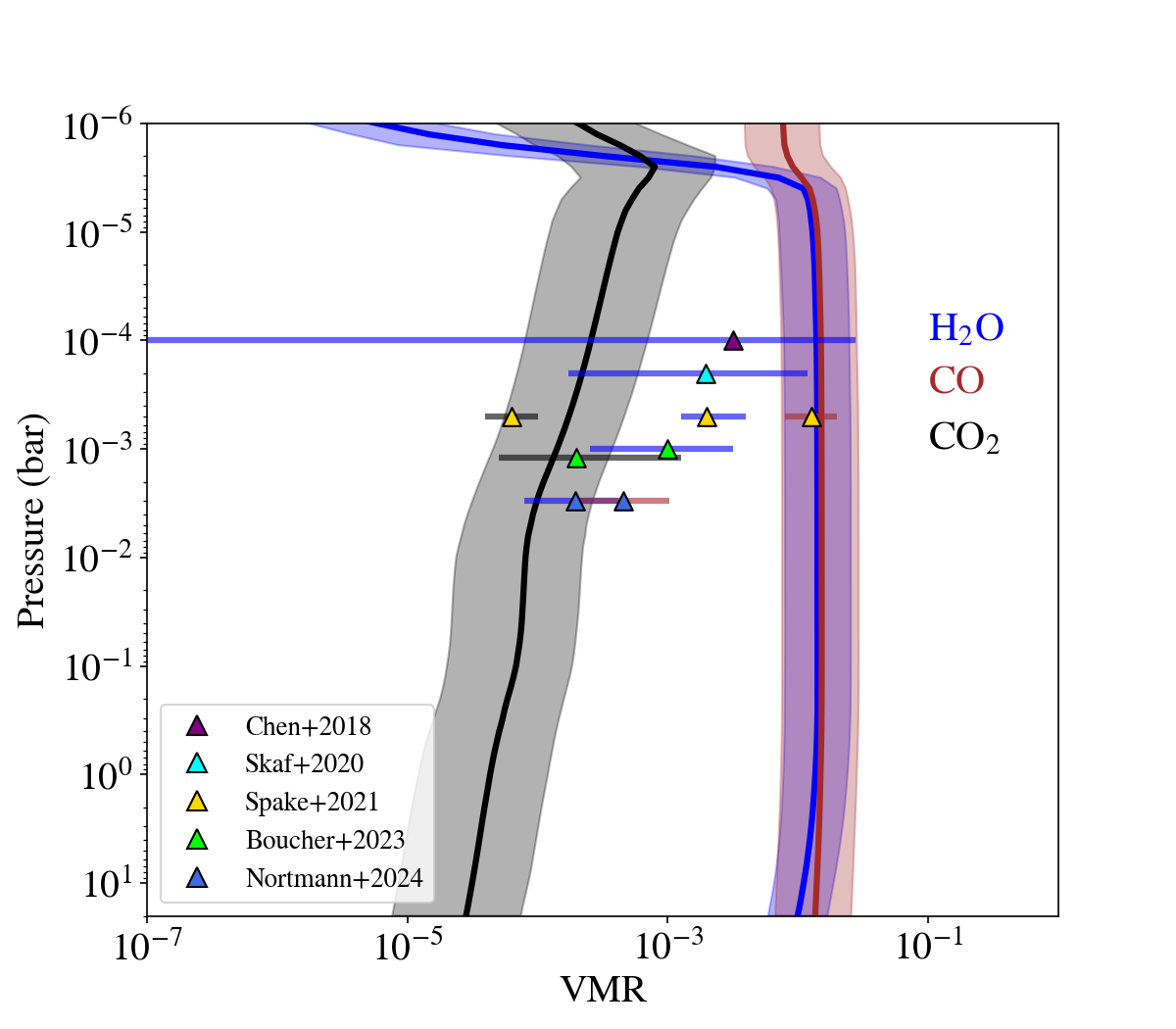}
\caption{Our abundance profiles of H$_2$O, CO, and CO$_2$ from the RCPE retrieval compared against the constraints obtained from previous studies (triangle markers). For the three molecules, lighter shades of their color represent the 1$\sigma$ confidence region. The error bars on the constraints of the abundances of H$_2$O, CO, and CO$_2$ are represented in blue, brown and black bars respectively. The constraints of \cite{spake2021} were taken from their estimate on WASP-127 b metallicity from their chemically consistent retrieval (i.e., 17 $\pm$ 4$\times$ solar). Our RCPE results agree well with the results reporting super-solar metal enrichment for WASP-127 b.
\label{fig:appendix-vmrs-comparision}}  
\end{figure}

Regardless of the model assumptions in all these studies, there is a general consensus that the atmosphere of WASP-127 b is metal-enriched, in accordance with the mass-metallicity relationship of exoplanets as discussed in \cite{spake2021}. From our RCPE and free-chemistry retrievals, we achieve a similar conclusion. More importantly, the abundance of CO$_2$ reported by previous studies can only be reproduced with our RCPE grid retrieval results. In Figure \ref{fig:appendix-vmrs-comparision}, we have shown that our RCPE results are in agreement with several previous studies that report super-solar metal enrichment in the atmosphere of WASP-127 b.

Our super-solar metallicity constraint ($\sim$  39$\times$ solar) from the photochemical retrieval (1D-RCPE) is well below the maximum metallicity for WASP-127 b ($\sim$ 600$\times$) based on interior structure modelling \citep[e.g.,][]{thorngren-fortney-2019}. While it might seem that such metal enrichment poses a challenge in explaining the low density of WASP-127 b, the resulting enhanced opacity could be a possible driver of inflation by delaying the atmospheric contraction due to heat deposition \citep{burrows2007}. The obtained high-altitude cloud deck further supports this reasoning. However, enhanced opacity is one of the many mechanisms considered to explain the inflated states of planets such as WASP-127 b e.g., Ohmic heating \citep{ohmic_heat_thorngren_fortney_2018, ohmic_heat_bonan_pu}, semi-convective interior \citep{layered_convection_heat_chabrier_2007}, He-rain \citep{helium-rain_heat_radius_inflation}. Disentangling these drivers of inflation is beyond the scope of our study and warrants further investigation.


\subsection{Validating the depletion of H$_{2}$S}

To confirm or rule out the preliminary signs of photochemistry we reported here, we strongly encourage additional observations and analyses on WASP-127 b. A possible venue to support/contradict our H$_{2}$S depletion would be by the detection/absence of SO$_2$. In the presence of strong UV irraadtion from the host star, SO$_2$ is an expected primary photochemical product from the oxidisation of H$_{2}$S \citep{zahnle2016_so2_hazes_eri, vulcan_tsai, polman_h2s_so2}. In IGRINS bandpass, SO$_2$ possess weak opacity and will be overshadowed by H$_2$O, CO opacities and their abundances. This statement also applies for CO$_2$ and could possibly be the reason why we do not detect it from our CC maps. NIRSpec and MIRI on the JWST are an excellent platform to provide an unambiguous SO$_2$ and CO$_2$ detections \citep{tsai2023_wasp39b_so2, co2_jwst, powell_so2_jwst, dyrek_so2_jwst} and with our analysis, we present an opportunity for further investigation.

\subsection{Rossiter-Maclaughlin effect and center-to-limb variations}

For our analysis, we did not model the Rossiter-Maclaughlin effect (RME) \citep{rossiter1924, maclaughlin1924, bro16, triaud2018, genest2022} and the center-to-limb variations (CLV) \citep{yan_clv_2017,allart2020,genest2022, sicilia_clv_2022}. CLVs arise from the non-uniformity of the stellar flux across the stellar disk from center to the outer periphery. In exoplanet spectroscopy where the spectra are derived using the flux ratio between the star and the planet, CLV is responsible for overestimation of calculated absorption depths of planetary spectral lines \citep{borsa_clv_rme_2018, sicilia_clv_2022}. However, the effect of CLV in the infra-red (IR) wavelengths is lesser than compared to visible region \citep{borsa_clv_rme_2018}. To assess this, we obtained the quadratic limb darkening (LD) coefficients of WASP-127 from \textsc{EXOFAST} \footnote{https://astroutils.astronomy.osu.edu/exofast/limbdark.shtml} and examined the LD multiple factor as a function of transit in H, K and V-bands. The variation in LD multiplicative factor value was less pronounced in the H and K bands, where the value changes from 0.74 (ingress/egress) - 0.99 (mid-transit). However, this variation in the V-band was significant where the value changed from 0.32 (ingress/egress) - 0.99 (mid-transit). Therefore, we expect the CLV effect on our results to be negligible. However, a detailed exploration of CLV effects on high-resolution IR data warrants further investigation. 

On the other hand, modelling the RME is imperative because the trace of stellar lines is left on the post-SVD data cube when there are common spectral lines for the planet and the star. While H$_2$O line are not expected to be present in a WASP-127 (G5-type), CO lines would bias the abundance constraints of WASP-127 b if not properly accounted for. However, with a slow rotation velocity of WASP-127 (\textit{vsini} $\sim$ 0.5 km/sec), \cite{allart2020} showed the RME on the transmission spectra on WASP-127 b was always encompassed inside the stellar line cores, thereby having minimal effect on the planet's spectrum. 

\section{Conclusions}
\label{sec:conclusions}
In this work, we analysed the time-resolved IGRINS high-resolution transmission spectra of WASP-127 b. We used the cross-correlation technique with a solar RCTE model containing 10 dominant IR absorbers to detect the atmospheric trail of WASP-127 b. From the CC maps, we detect a strong signal of H$_2$O (8.67$\sigma$), supporting its previous detections from HST \citep{spake2021} and SPIRou \citep{boucher2023}. We masked the strong telluric wavelengths and cropped the out-of-transit phases to reveal a strong CO signal (4.01$\sigma$) from the CC maps. Among these, cropping the out-of-transit frame reduced the CC noise and provided a significant boost to the CO signal. We further validated this detection by selecting the dominant orders with CO opacity to produce a CC map with an amplified SNR (4.34$\sigma$) of the CO signal.



We assessed the molecular abundances and temperature structure of WASP-127 b's atmosphere using a free-chemistry retrieval, a RCTE grid retrieval, and a RCPE grid retrieval. From the free-chemistry retrieval, we obtained super-solar abundance limits (within 2 $\sigma$) for both log$_{10}$$X_\mathrm{H_2O}$ (--1.23$^{+0.29}_{-0.49}$) and log$_{10}$$X_\mathrm{CO}$ ($>$--2.20). For the rest other atmospheric constituents, we place upper limits on their abundances. We also retrieved the pressure level for a high-altitude cloud deck at $\sim$ 0.06 mbar (log$_{10}$$ P_{cl}$ = --4.21$^{+0.48}_{-0.36}$). 

From the chemically consistent RCTE grid retrieval, we obtained an upper limit on the [M/H] as $<$ 1.64 (2$\sigma$) and a bounded constraint on the sub-solar C/O as 0.34$^{+0.08}_{-0.09}$. Conversely, we obtained a bounded contraint on the [M/H] as 1.59$^{+0.30}_{-0.30}$ and an upper limit on the C/O as $<$ 0.68 from the RCPE grid retrieval. We obtained a cloud pressure level (log$_{10}$$P_{cl}$ = --3.73$^{+0.27}_{-0.27}$) from the RCPE grid retrieval, which is consistent with the results of \cite{Skaf_FeH_HST} $\&$ \cite{allart2020} within 1$\sigma$. However, a deeper cloud deck was obtained at $\sim$ 3.4 mbar (log$_{10}P_{cl}$ = --2.46$^{+0.41}_{-0.69}$) from the thermochemical counterpart. The heat redistribution factor ($f$) remained the only parameter consistent among the two grid retrievals rendering a similar atmospheric temperature structure from the two grid retrievals. 

We attributed this discrepancy on [M/H] and C/O to the mandatory presence of H$_2$S from a thermochemical atmospheric assumption. Our test grid retrievals, where we modified the H$_{2}$S abundance profile with a scaling factor, indicated that the RCTE retrieval heavily relied on the presence of H$_2$S. However, we did not find any trace of this molecule from our CC analysis and the free-chemistry retrieval. Due to the H$_{2}$S depletion in the upper atmosphere, modifying the H$_{2}$S abundance did not affect our inferences from the RCPE retrieval. Furthermore, the Bayesian evidence strongly supports that a photochemical model performed better in matching the data when compared to a thermochemical model. Based on all these indicators, we draw our main conclusions from the RCPE/photochemical grid retrieval that the atmosphere of WASP-127 b is super-solar ($\sim$ 39$\times$) with C/O ratio $<$0.68.

A tighter constraint on the C/O ratio requires additional observations covering the dominant wavelength regions of CO$_2$ opacity. In the wavelength coverage of IGRINS, opacity from CO$_2$ is overshadowed by H$_2$O and CO. Given recent detections of CO$_2$ with JWST \citep{co2_jwst,  bean_2023_nature_hd149, qiao_jwst_hd209},  we anticipate that the observations with NIRSpec and MIRI should not only tighten our C/O inference, but also validate the evidence of photochemistry presented here.

\begin{acknowledgments}
This work used the Immersion Grating Infrared Spectrometer (IGRINS) that was developed under a collaboration between the University of Texas at Austin and the Korea Astronomy and Space Science Institute (KASI) with the financial support of the Mt. Cuba Astronomical Foundation, of the US National Science Foundation under grants AST-1229522 and AST1702267, of the McDonald Observatory of the University of Texas at Austin, of the Korean GMT Project of KASI, and Gemini Observatory. K.K., M.R.L., and J.L.B. acknowledge support from NASA XRP grant 80NSSC19K0293. M.R.L. and J.L.B. acknowledge support for this work from NSF grant AST-2307177. L.W. and M.W.M. acknowledge support from NASA Hubble Fellowship program.  We would also like to thank the NOIRLabs support staff helping with the implementation of these observations. Support for this work was provided by NASA through the NASA Hubble Fellowship grants HST-HF2-51485.001-A and HST-HF2-51496.001-A awarded by the Space Telescope Science Institute, which is operated by AURA, Inc., for NASA, under contract NAS5-26555. Finally, we acknowledge the Research Computing at Arizona State University for providing HPC and storage resources that have significantly contributed to the research results reported within this manuscript.
\end{acknowledgments}

%

\vspace{5mm}
\facilities{Gemini South (IGRINS)}


\software{\textsc{pymultinest} \citep{Buchner2014}, \textsc{astropy} \citep{astropy2022}, \textsc{Numpy} \citep{numpy}, \textsc{Scipy} \citep{scipy}, \textsc{Matplotlib} \citep{matplotlib}}



\appendix
\label{sec:appendix-add-figures}

\begin{figure}[ht!]
\centering
\includegraphics[scale = 0.3]{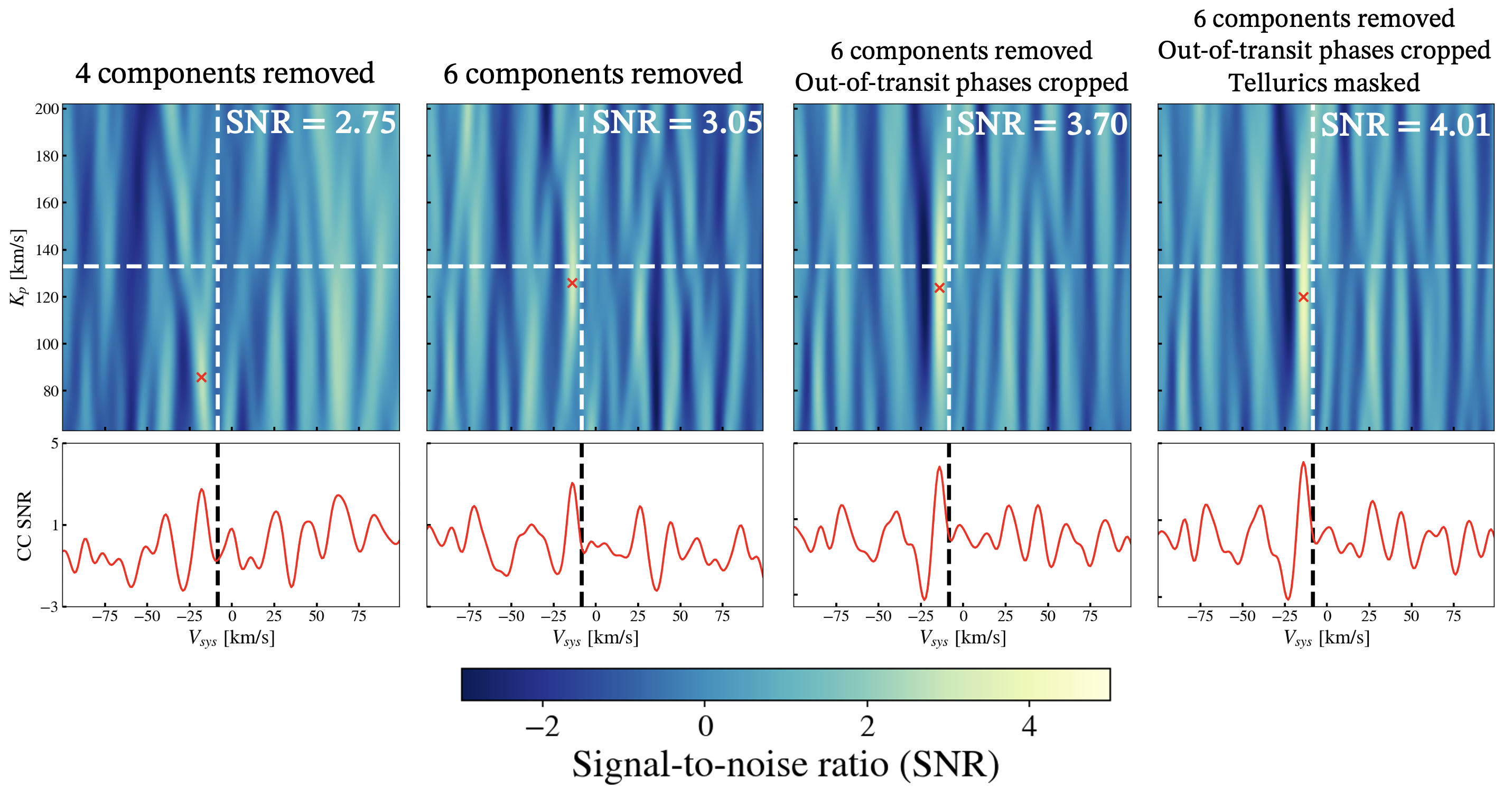}
\caption{The steps implemented on the data of WASP-127 b to highlight the weak CO signal. From the leftmost panel, it is not possible to claim a confident detection of CO. The most significant increase in the significance of our CO detection comes by cropping the out-of-transit phases post-SVD. 
\label{fig:appendix-flowchart}}
\end{figure}

\begin{figure*}[ht!]
\begin{center}
\includegraphics[scale = 0.25]{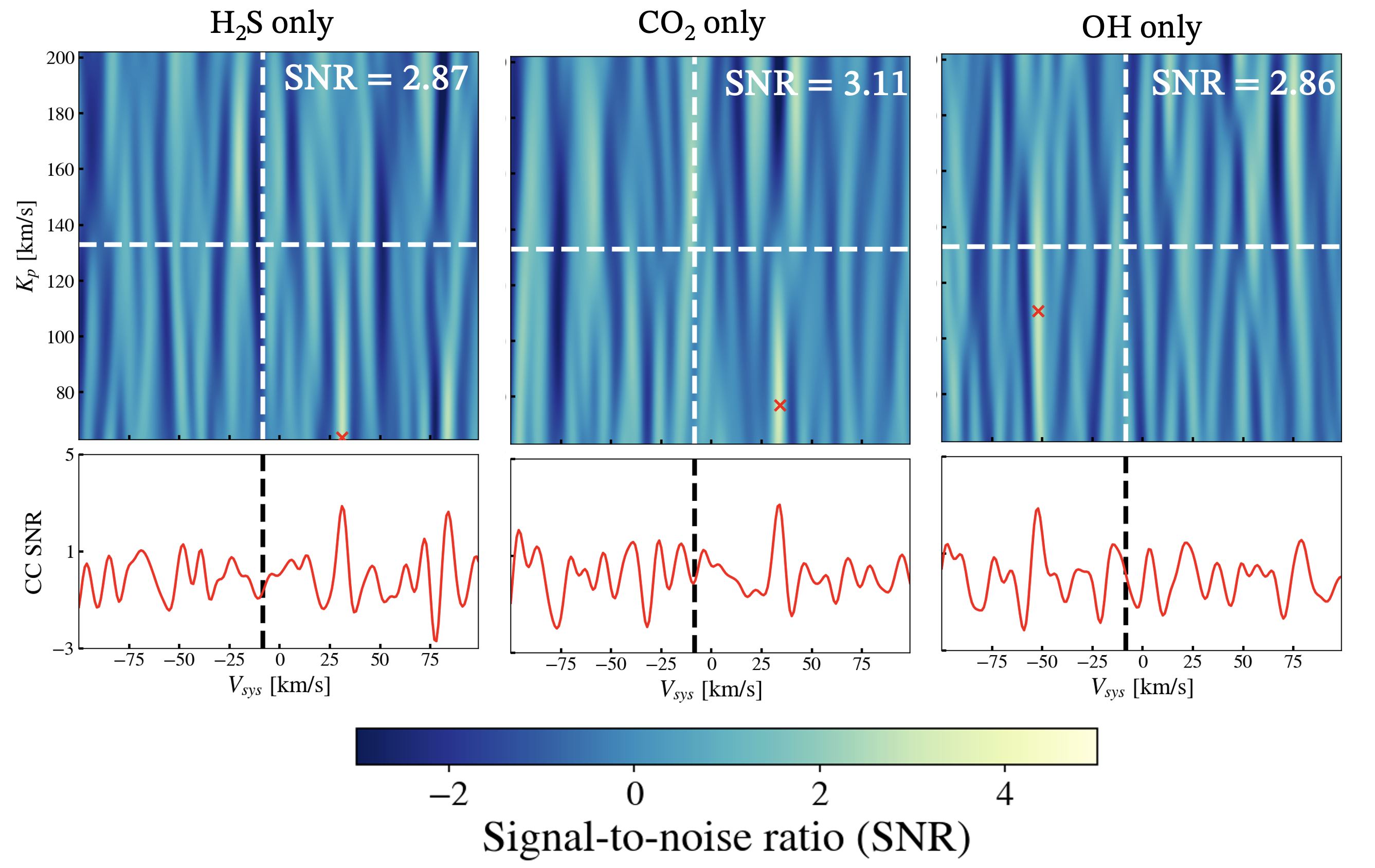}
\caption{The CC maps showing the of non-detections of CO$_2$, H$_2$S, and OH from our dataset. These three maps are a subset of eight non-detection maps of the absorbers included in our models apart from H$_2$O and CO. 
\label{fig:appendix-non-detections}}  
\end{center}
\end{figure*}

\begin{figure*}
    \centering
    \includegraphics[width = \textwidth]{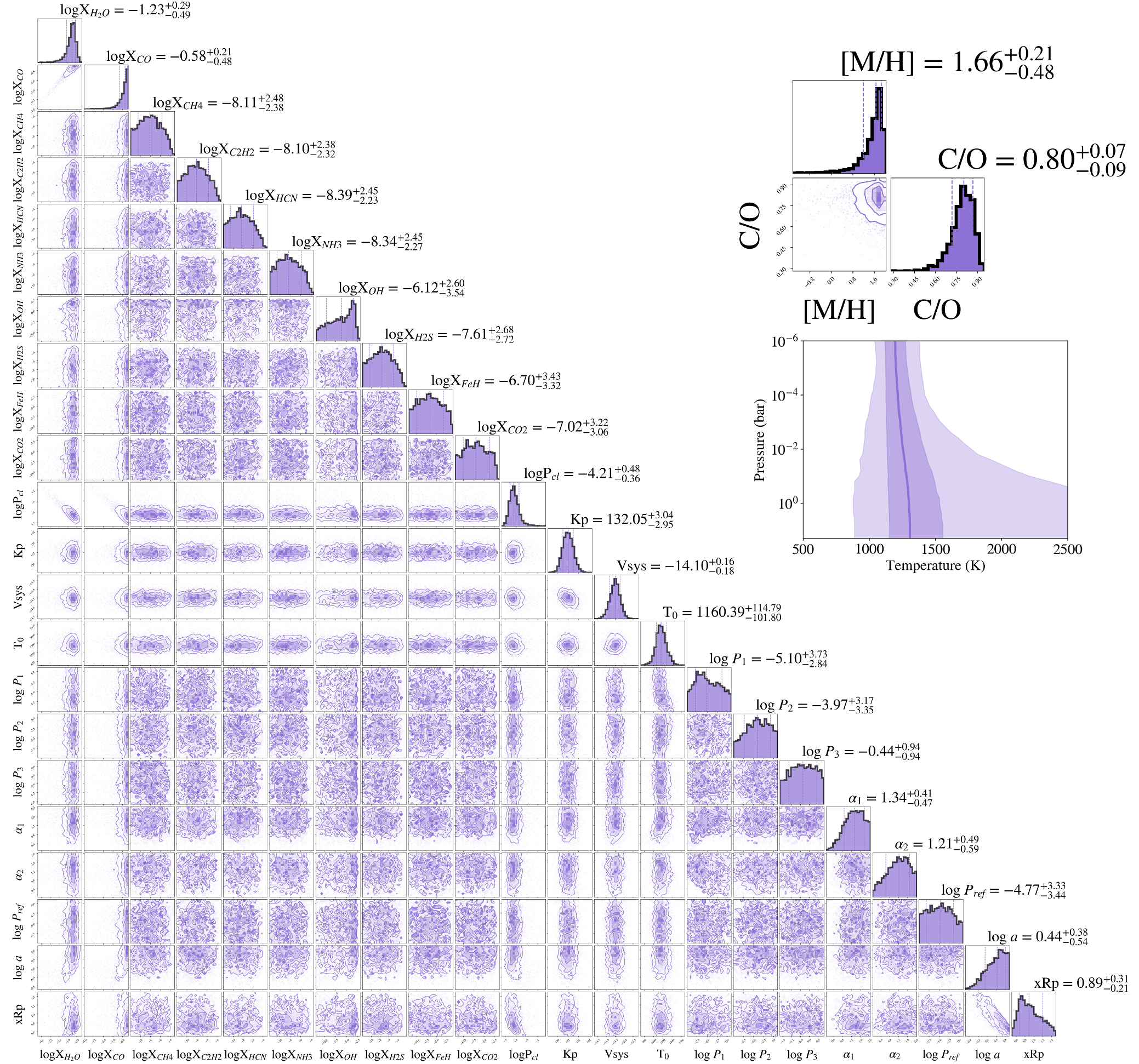}
    \caption{The full posterior from the free-chemistry retrieval showing the constraints on all the parameters specific to free-chemistry. In all the marginal distributions, the median and the 1$\sigma$ bound are shown as the vertical dashed lines. The posteriors on [M/H] and C/O are calculated based on the obtained abundance constraints (see \cite{brogi23} for the description). The obtained temperature structure based on the six parameters of the TP profiles is shown in the top right. 
    \label{fig:appendix-full-free-retrieval}}
\end{figure*}

\begin{figure*}[ht!]
\begin{center}

\includegraphics[scale = 0.33]{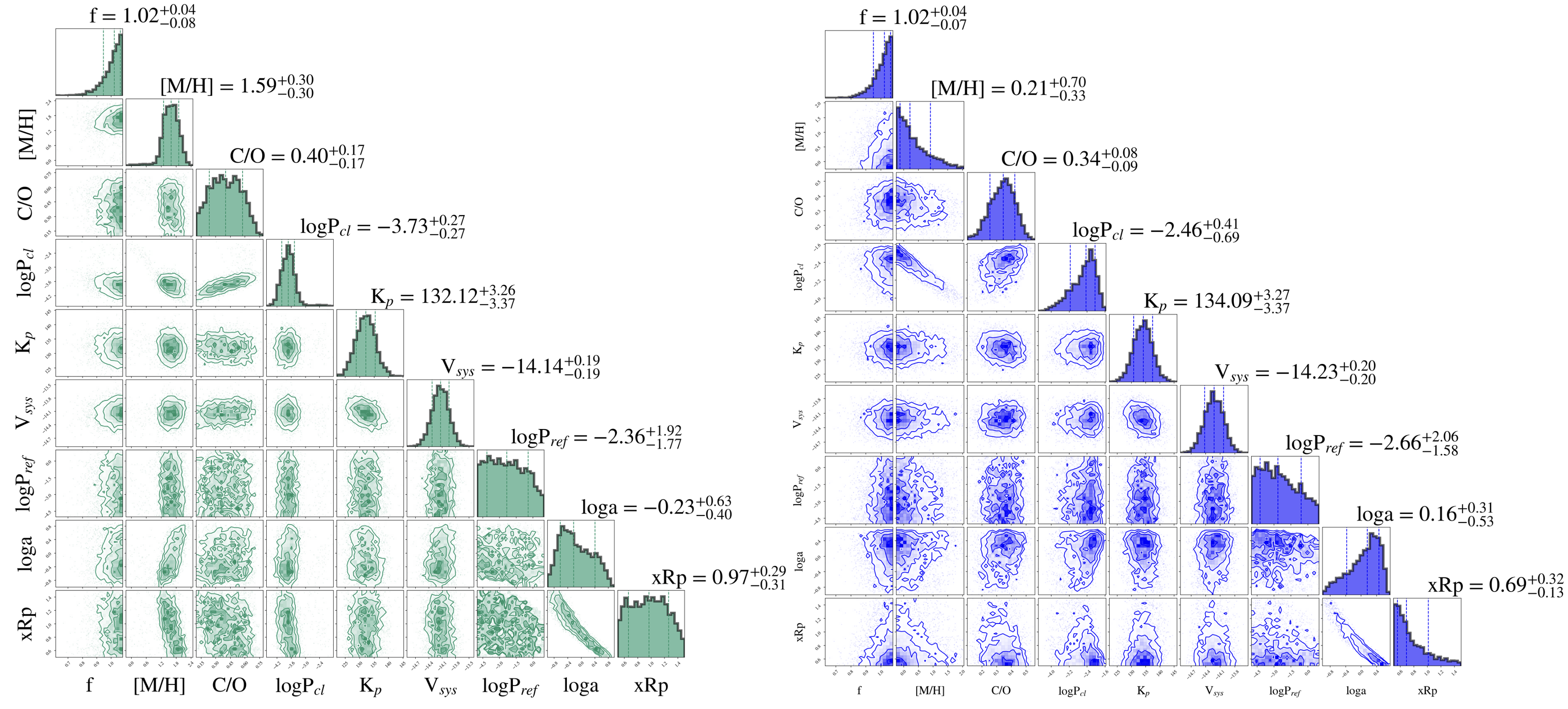}
\caption{The full posteriors of the RCTE (right, blue) and the RCPE (left, green) grid retrievals. The main conclusions in this manuscript have been drawn from the RCPE grid retrieval. In all the marginal distributions, the median and the 1$\sigma$ bound are shown with a vertical dashed line and their values are given in the top of each column.
\label{fig:appendix-both-grid-retrievals}}  
\end{center}
\end{figure*}

\begin{figure*}
\begin{center}
\includegraphics[scale = 0.33]{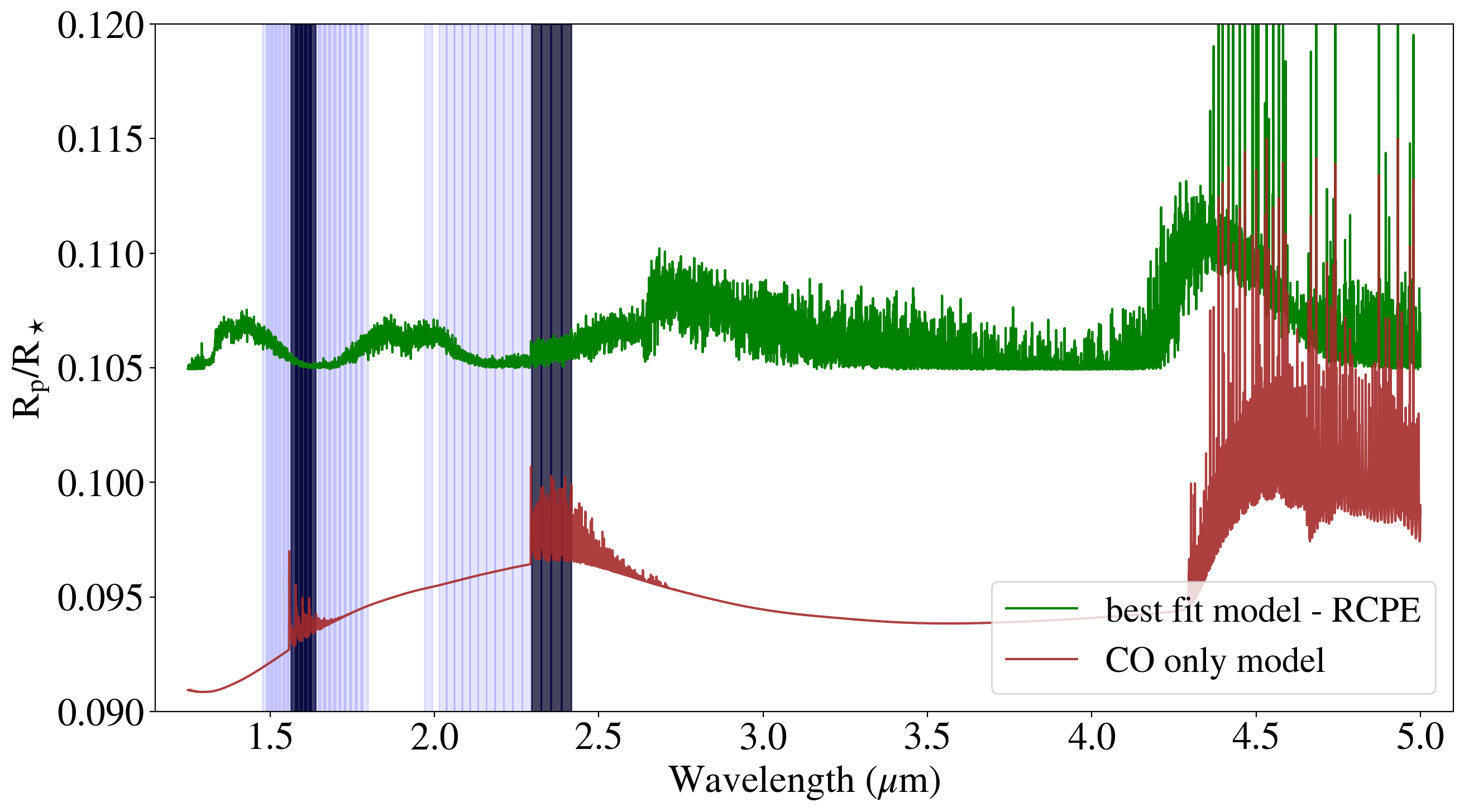}
\caption{The solar RCTE model of CO (brown) used for our cross-correlation analysis. The best fit RCPE model is also shown as the green curve. The 38 IGRINS spectral orders used in our analysis are shown as the blue stripes. Among these, the eight spectral orders where the CO opacity is dominant as shown in black. The CC map where we detected maximum SNR of CO signal has been generated with the CC values from these eight spectral orders.
\label{fig:appendix-order_selection_models}}
\end{center}
\end{figure*}
\clearpage

\begin{figure*}
\centering
\includegraphics[scale = 0.28]
{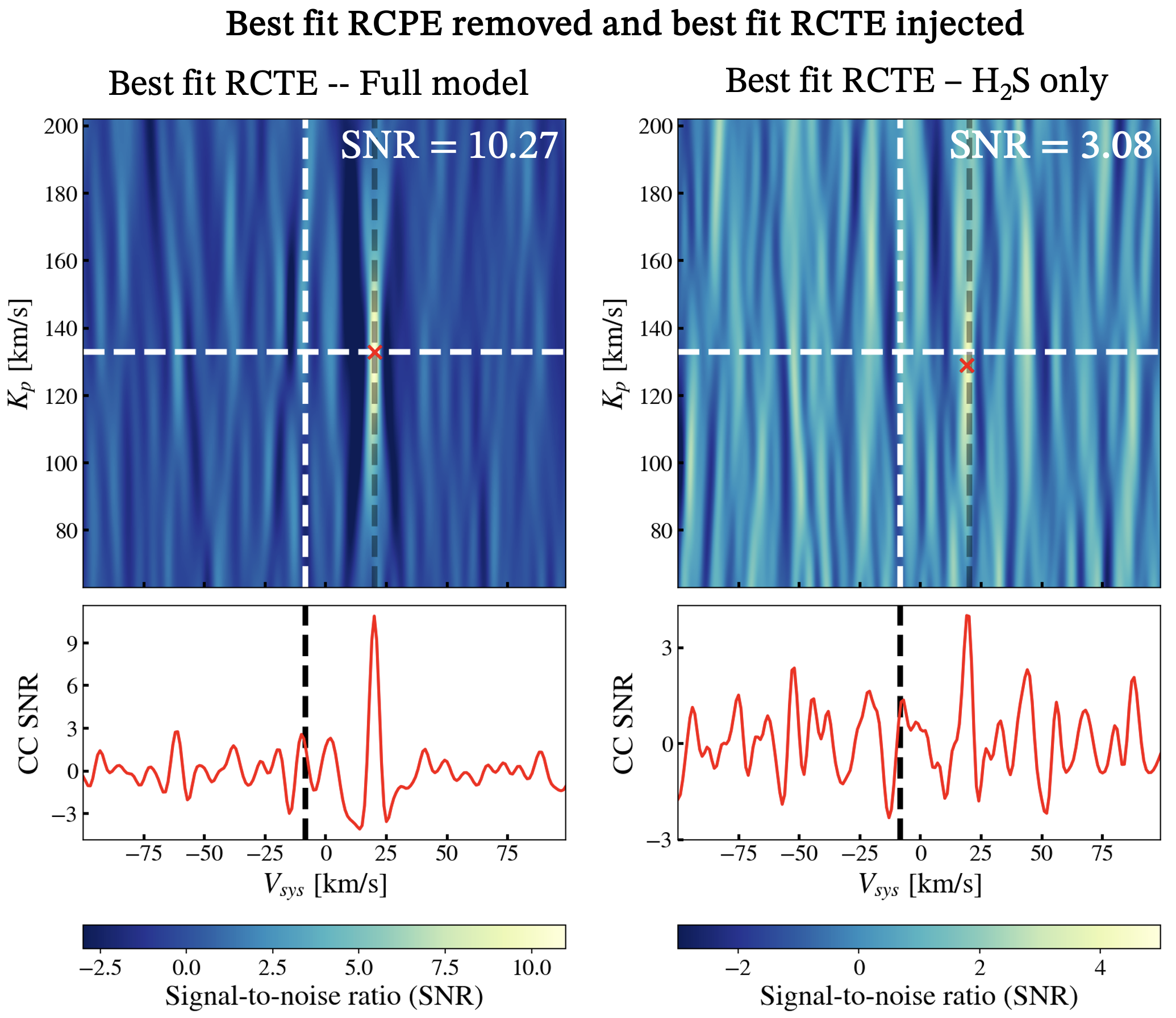}
\caption{The CC maps where were removed the best fit RCPE model from the data and injected the best fit RCTE model at V$_{sys}$ = 20 km/sec (Black dashed vertical line in the maps). The cross-correlation was performed with best fit RCTE full model and H$_2$S absorption only model to assess the level at which H$_2$S can be detected, if present, dictated by thermochemical equilibrium. While not as strong as the full model (10.27$\sigma$), we would have been able to detect H$_2$S at 3.08$\sigma$ under thermochemical equilibrium.  
\label{fig:appendix-h2s_injection_recovery}}  
\end{figure*}

\clearpage

\bibliography{sample631}{}
\bibliographystyle{aasjournal}



\end{document}